\documentclass{aastex62}

\graphicspath{{./}{figures/}}

\received{January 1, 2018}
\revised{January 7, 2018}
\accepted{\today}
\submitjournal{ApJ}

\shorttitle{Dark Matter in Galaxies Within $5r_e$}
\shortauthors{W. Harris et al.}
\usepackage{longtable}
\usepackage{threeparttable}

\begin{document}

\title{Measuring Dark Matter in Galaxies:  The Mass Fraction Within 5 Effective Radii}

\correspondingauthor{William E. Harris}
\email{harris@physics.mcmaster.ca}
\author[0000-0001-8762-5772]{William E. Harris}
\affil{Department of Physics and Astronomy, McMaster University, Hamilton, ON L8S 4M1, Canada}

\author{Rhea-Silvia Remus}
\affil{Universit\"ats-Sternwarte M\"unchen, Scheinerstra{\ss}e 1, D-81679 M\"unchen, Germany}

\author[0000-0002-9451-148X]{Gretchen L. H. Harris}
\affil{
Department of Physics and Astronomy, Waterloo Institute for Astrophysics, University of Waterloo, Waterloo, ON N2L 3G1, Canada}

\author{Iu. V. Babyk}
\affil{Department of Physics and Astronomy, University of California Irvine, 4329 Frederick Reines Hall, Irvine, CA 92697-4575, USA}
\affil{Main Astronomical Observatory of the National Academy of Sciences of Ukraine, 27 Zabolotnoho str., Kyiv 03143, Ukraine}

%\nocollaboration
%\nocollaboration

%\nocollaboration

\begin{abstract}

Large galaxies may contain an ``atmosphere'' of hot interstellar X-ray gas, and the temperature
and radial density profile of this gas can be used to measure the total mass of the
galaxy contained within a given radius $r$.  We use this technique for 102 early-type
galaxies (ETGs) with stellar masses $M_{\star} > 10^{10} M_{\odot}$, 
to evaluate the mass fraction of dark matter (DM) 
within the fiducial radius $r = 5 r_e$, denoted $f_5 = f_{DM}(5r_e)$.
On average, these systems have a median $\overline {f_5} \simeq 0.8 - 0.9$ 
with a typical galaxy-to-galaxy scatter
$\pm 0.15$.  Comparisons with mass estimates made through the alternative techniques of 
satellite dynamics (e.g. velocity distributions of globular clusters, planetary nebulae, 
satellite dwarfs) as well as strong lensing show encouraging consistency over the same range
of stellar mass.  We find that many of the disk galaxies (S0/SA0/SB0) 
have a significantly higher mean $f_5$ than do the pure ellipticals, 
by $\Delta f_5 \simeq 0.1$. We suggest that this higher level may be a consequence
of sparse stellar haloes and quieter histories with fewer major episodes of feedback or mergers.
Comparisons are made with the Magneticum Pathfinder suite of simulations for both
normal and centrally dominant ``Brightest Cluster'' galaxies.  Though the observed data exhibit somewhat larger
scatter at a given galaxy mass than do the simulations, the mean level of 
DM mass fraction for all classes of galaxies is in good first-order agreement with the simulations.  
Lastly, we find that the group
galaxies with stellar masses near $M_{\star} \sim 10^{11} M_{\odot}$ have
relatively more outliers at low $f_5$ than in other mass ranges, 
possibly the result of especially effective AGN feedback in that mass range leading to 
expansion of their dark matter halos.
\end{abstract}

%% Keywords should appear after the \end{abstract} command. 
%% See the online documentation for the full list of available subject
%% keywords and the rules for their use.

%% From the front matter, we move on to the body of the paper.
%% Sections are demarcated by \section and \subsection, respectively.
%% Observe the use of the LaTeX \label
%% command after the \subsection to give a symbolic KEY to the
%% subsection for cross-referencing in a \ref command.
%% You can use LaTeX's \ref and \label commands to keep track of
%% cross-references to sections, equations, tables, and figures.
%% That way, if you change the order of any elements, LaTeX will
%% automatically renumber them.
%%
%% We recommend that authors also use the natbib \citep
%% and \citet commands to identify citations.  The citations are
%% tied to the reference list via symbolic KEYs. The KEY corresponds
%% to the KEY in the \bibitem in the reference list below. 

\section{Introduction}\label{sec:intro}

The total mass and mass profile of a galaxy are fundamental tracers of
its evolutionary history.  But since most of the mass of a galaxy is in the form
of dark matter (DM), the mass profile must be determined indirectly through
the use of visible tracers of various kinds.

The DM fraction of mass within a given radius $r$ is simply 
\begin{equation}
    f_{DM}(r) = 1 - {M_{bary}(r) \over M_{tot}(r) }
\end{equation}
where $M_{bary}$ and $M_{tot}$ denote the total baryonic mass and total gravitating mass 
within $r$. Since in most galaxies the stellar mass is more centrally concentrated
than the DM halo, $f_{DM}$ should increase with $r$ 
\citep[e.g.][]{deason_etal2012,tortora+2014,alabi_etal2016,alabi_etal2017}. 
But in addition, recent theory indicates that the ratio of baryonic mass 
to DM within a given radius should also depend on galaxy total mass, environment, and evolutionary history
including the epochs and amounts of gas infall, merging, and feedback 
\citep[cf. the references cited above, as well as][among others]{remus+2017,lovell+2018,wojtak_mamon2013,hirschmann:2014,elias+2018,monachesi+2019,d'souza+2018,tortora+2019}.
Observational measurements of $f_{DM}$ can therefore provide markers of those histories.

The radius $r$ is often normalized in units of the effective radius $r_e$ of
the stellar light that encloses half the (projected 2D) total luminosity; 
most recent discussions (see the papers cited above) have tended to focus 
on $f_{DM}$ within fiducial radii of either $1 r_e$ or $5 r_e$.  
The radius of $5 r_e$ encloses a large volume well outside the 
merger and star-forming activity often contained within the bulge and
inner halo (roughly, $r \lesssim 2r_e$).  At radii as large as $5 r_e$ and
beyond, the recent models and simulations indicate that we should expect
a relatively high mean DM fraction, but perhaps with outliers at lower $f_{DM}$ that
preserve the record of 
the most major merger, feedback, or accretion events
\citep[cf.][]{deason_etal2012,wojtak_mamon2013,remus+2017,forbes+2017,lovell+2018}.
%For late-type galaxies (LTGs) that are rotation dominated, a rough analog is the fiducial radius $2.2R_d$
%where the peak of the rotation curve is reached for simple exponential disks with no DM
%and where $R_d$ is the disk scale length \citep[e.g.][]{courteau+2015}.

Comparisons between theoretical models and observational data can now be done through high-resolution
simulations of galaxy formation rather than simpler analytical models; and on the observational
side, the amount and quality of measured mass profiles is steadily increasing.
Direct measurement of $M(r)$ for early-type galaxies (ETGs) has most often been done
through the radial velocity distributions of satellites such as old halo stars, globular clusters (GCs),
planetary nebulae (PNe), or dwarf satellite galaxies.  For the Milky Way, tangential velocities (from proper motions)
can be added to the mix, enabling narrower constraints on the phase-space distributions of
its satellites.  Many analytical methods have been developed
for external galaxies, including the Projected Mass Estimator (PME), the Tracer Mass Estimator (TME),
solutions of the Jeans equations, orbit libraries and made-to-measure codes, or the distribution function in phase space
\citep[cf.][among others]{rix+1997,wu_tremaine2006,deason_etal2012,watkins+2010,delorenzi+2007,napolitano+2009,romanowsky+2009,cappellari+2013,courteau+2014,alabi_etal2017,eadie+2019}.  
Well known uncertainties affecting these methods 
    include the anisotropy of the tracer orbital distributions (since only the radial velocities of the tracers are measured), the slope of
    the gravitational potential, and the presence of substructure amongst
    the tracers, all of which can
    differ strongly and unpredictably from one target galaxy to another.  
    In addition, at larger radii
    the estimated $M(r)$ may depend heavily on small numbers of tracers that
    lie at the largest observed radii.  These issues and others are discussed at length 
    by \citet{alabi_etal2016, alabi_etal2017} (hereafter denoted A16, A17) and in 
    the other papers cited above.
    
High-mass galaxies (Milky-Way-sized and above) may also hold
significant amounts of diffuse, hot interstellar X-ray gas.  Using the temperature and density
distribution of this gas opens up an entirely different approach for
deducing the mass profile of its host galaxy \citep[e.g.][among many others]{bahcall_sarazin1977,fabricant_gorenstein1983,nulsen_bohringer1995,irwin_sarazin1996,brighenti+1997,loewenstein_mushotzky2003,fukazawa+2006,babyk_etal2018,harris+2019}.  
Such work has added substantial evidence for the presence
of DM at scales ranging from individual galaxies out to their larger groups and clusters.
There are now enough individual galaxies with X-ray studies in the literature to permit a new  
look at the pattern of DM mass fraction at radii reaching the outer halo.

Both the X-ray gas method and the 
tracer satellite method have uncertainties
and potential biases for deriving $M(r)$ (which represents 
the depth of the potential well that ultimately drives both the gas temperature
and the amplitude of the satellite motions).  These will be discussed further below. Perhaps the most important point to highlight, however, is that the two methods have 
\emph{different} inbuilt biases and uncertainties and  
are encouragingly close to being physically independent. 
Comparisons between them should therefore be worthwhile.
Direct comparisons of the estimates of $M(r)$ from satellite dynamics  
and X-ray gas have been done for only a handful of 
relatively nearby, giant galaxies \citep[e.g.][among others]{cohen_ryzhov1997,cote+2003,bridges+2006,schuberth+2006,romanowsky+2009,longobardi+2018}, all of which have
rich GC and PNe populations.  These detailed individual studies are extremely valuable. 
However, detection and characterization of large-scale trends of DM fraction with
galaxy mass and morphology, and followup comparison with galaxy evolution
modelling, needs much larger observational samples. 

Our present paper has two primary goals:  (1) We compare measurements
of the DM mass fraction $f_{DM}$ obtained by the X-ray gas method 
with two other very different methods:  satellite dynamics, and strong lensing, using 
previously published data from the recent literature.
(2) We assess how well these current sets of data 
agree with one particular suite of theoretical realizations for galaxy formation,
the recent Magneticum Pathfinder simulations from which $f_{DM}$ can be predicted \citep[i.e.][]{remus+2017}.  Our findings are that there is now excellent first-order 
concordance among these simulations and the different observational methods, but
interesting differences in detail show up that may be connected with the
evolutionary histories as well as features of the simulations.

The outline of this paper is as follows.  In Section 2  we provide background on the data for the
various mass parameters;  Section 3 shows the resulting $f_{DM}$ 
distributions versus galaxy mass; and Section 4 gives an overall discussion and comparison with selected
model simulations.  In Sections 5, 6, and 7 we provide an overview, prospects for
the next steps in this investigation, and a summary.

A distance scale $H_0 = 70$ km s$^{-1}$ Mpc$^{-1}$ is assumed throughout this discussion.  
For convenience, in what follows we denote $f_{DM}(5 r_e)$ more concisely as $f_5$. 
We also denote $M_{tot}(5 r_e)$, the total gravitating mass within $5 r_e$, simply as $M_5$.
We will also refer to the enclosed masses derived from either the X-ray gas profiles or
the dynamics of tracer objects (GCs, PNe, dwarf satellites) as simply
the ``X-ray'' and ``Satellite'' masses in the various figures and discussion
to follow.

\section{The Data}\label{sec:data}

In \citet{harris+2019} (hereafter H19), we discussed a
subset of 45 galaxies for which information about both their
X-ray atmospheres and 
their GC populations is available.  These 45 comprise the galaxies that appear in both
the GC system catalog of \citet{hha2013} and the X-ray list of \citet{babyk_etal2018}
(hereafter B18).
H19 derived correlations among gas mass and total gravitating mass within $5 r_e$, 
total stellar mass $M_{\star}$, GC
system mass $M_{GCS}$, and total halo (virial) mass $M_h$, and finally the
correlations of $f_5$ with both $M_{\star}$ and $M_h$.  We found that almost 
90\% of this restricted sample fell along a consistent mean level
of $\langle f_5 \rangle = 0.83$ with a dispersion of only $\sigma(f_5) = 0.07$
and a handful of outliers falling below $f_{DM} \lesssim 0.6$.  This pattern,
though still sketchy, proved
to be strikingly similar to predictions from two recent hydrodynamical simulations,
specifically the Magneticum Pathfinder suite \citep{remus+2017}, and the Illustris TNG suite \citep{lovell+2018}.

The observational correlation of $f_5$
with $M_{\star}$ has recently been analyzed 
as well by A16 and A17 from their velocity measurements for GCs,
combined with PNe velocities from the previous literature, as satellite tracers.
In their results, 32 individual galaxies yielded $f_5$ values that 
spread across almost the entire physically permitted range, from 
$f_5 \lesssim 0.1$ up to nearly 1.0.  Although 2/3 of these
fall in the range $f_5 \sim 0.6 - 0.9$, the remaining 1/3 scatter to much lower values 
and no clear systematic trend with total stellar mass is seen 
(see particularly Fig.~2 of A17).  Clearly, analyses of larger samples of
galaxies are desirable.

In the present paper, we drop any restrictions on comparisons with GC/PNe satellite 
populations and concentrate on results from X-ray data alone.
B18 provide a homogeneous set of measurements of the total X-ray radial profiles, 
the gas mass $M_X$, and $M_5$ for 94 relatively
massive galaxies nearer than $\sim 200$ Mpc.  The great majority of
these are ETGs (ellipticals or S0 disk galaxies), but the sample also includes a
few late-type galaxies (LTGs) that happen to have measurable amounts of X-ray gas.  
They cover the full range of galaxy
environments, from relatively isolated systems up to BCGs 
(Brightest Cluster Galaxies) and BGGs (Brightest Group Galaxies)
at the centers of clusters; we will refer to those giants as ``centrals'' and
the other galaxies as ``normals''.  
From this list of 94, we have deleted 16 with the most uncertain
measurements (see below), leaving 78 systems.  We have, however, added
24 more ETGs with \emph{Chandra} data newly measured through exactly the same
procedures by Babyk (2020, in preparation), making a final
total of 102 galaxies with measured mass distributions based on their X-ray
gas content.  This sample is significantly larger than the one in H19 and covers
a wider mass range.

Basic parameters for this target list are given in Table \ref{tab1}, including
the galaxy identification; group or cluster environment; Hubble type; de Vaucouleurs T-type; location on the sky
(RA, Dec for J2000); foreground extinction $A_B$; distance D; and effective
radius $r_e$.  These parameters are drawn from the HyperLeda catalog except for
the environments and foreground extinctions, which are taken from NED
(NASA Extragalactic Database).  In cases where no entry is given for the
environment, the galaxy is relatively isolated or part of a very small group. For the 
effective radii $r_e$, we used optical Digitized Sky Survey (DSS) images for our own
measurements, as noted in B18. We extracted $10' \times 10'$ images of each galaxy and
determined surface brightness profiles centered on the peak of the optical emission
through a curve-of-growth technique.  
We obtained the background level at large radius by fitting a constant to the brightness profile, 
and performed numerical integration to define the total optical flux as the emission above 
background by 5$\sigma$, and finally determined the uncertainties on $r_e$ by running 
1,000 Monte Carlo realizations.
%The effective radii $r_e$ are drawn from B18 and Babyk (2020).  

\begin{table*}[t]
\begin{center}
\caption{Basic Data for Target Galaxies\label{tab1}}
\begin{tabular}{lllccclcc}
\tableline\tableline\\
Name & Environment & Type & T & RA & Dec & $A_B$ & $D$ & $r_e$ \\
 & & & & (deg) & (deg) & (mag) & (Mpc) & (kpc) \\
 \\[2mm] \tableline\\
NGC193     & & SAB0  &  -0.3 &     9.827374 &     3.331333 &   0.10 &   63.5 & $ 11.4\pm  0.8$ \\
NGC315     & Zw0107.5+3212 & E     &  -4.1 &    14.453676 &    30.352354 &   0.28 &   73.6 & $ 17.6\pm  3.8$ \\
NGC326     & Zw0056.9+2636 & E     &  -0.4 &    14.594292 &    26.866278 &   0.29 &  216.3 & $ 38.2\pm  6.6$ \\
NGC383     & BGG, Zw0107.5+3212 & S0    &  -2.9 &    16.853779 &    32.412663 &   0.31 &   74.8 & $  9.6\pm  1.0$ \\
NGC499     & & E5    &  -2.9 &    20.797979 &    33.460316 &   0.31 &   64.7 & $ 11.4\pm  1.6$ \\
NGC507     & BGG, N507 Group & S0    &  -3.3 &    20.916471 &    33.255764 &   0.27 &   73.0 & $ 21.4\pm  2.4$ \\
\\
\tableline
\end{tabular}
\end{center}
\begin{tablenotes}
\item {\sc Note:} Only the first few lines of the table are listed as a guide to form and content.
The complete table is given in the on-line version of the paper.
\end{tablenotes}
\vspace{4mm}
\end{table*}

\subsection{X-ray Measurements}

For the X-ray data, full discussions of the measurements and data reduction are given
in B18 and H19 and we provide only a brief summary here.  
\emph{Chandra} X-ray observations with $> 10$ ks exposures of the target galaxies 
were used to extract exposure- and background-corrected images in the 0.5-6.0 keV energy band. Point sources and other non-X-ray-gas features were
detected and then removed by applying the \texttt{wavedetect} routine. 

The radial profiles were fitted with a single
$\beta$-model \citep{caval,gorenstein+1978} yielding a gas density profile
\begin{equation}
    \rho_g(r) = \rho_0\left(1+\left(\frac{r}{r_c}\right)^2\right)^{-3\beta/2},
\end{equation}
where $\rho_0$ = 2.21$\mu m_p n_0$ is the central gas density that can be found 
from the emissivity profile \citep[e.g.][B18]{ettori}, $n_0$ is the central
number density, and $r_c$ is the core radius of the density profile.  
The index $\beta$ refers physically to the ratio of specific
energies of the stellar component to the gas (see the references cited above),
but acts as a free parameter for the fit. The hot-gas mass within radius $r$ 
results from integration of the gas density profile as 
\begin{equation} 
    M_X(r) = 4 \pi \rho_0 \int_0^r r^2 \left(1+\left(\frac{r}{r_c}\right)^2\right)^{-3\beta/2} dr.
\end{equation}
Finally, the total gravitational mass within $r$ is calculated from the condition
of hydrostatic equilibrium, 
\begin{equation}
    M_{tot}(r) = -\frac{k T r}{G\mu m_p} \left(\frac{d \ln{\rho_g}}{d\ln{r}}\right) \, .
\end{equation}
The implicit assumptions of hydrostatic equilibrium and isothermality are used throughout (see B18).
At large radii well outside the X-ray core $r_c$, 
the expression for total mass simplifies to $M_{tot}(r) \simeq 3 \beta k T r/G \mu m_p$
as long as the gas follows the $\beta-$model density profile.

In all cases, the gas component as we use it here refers to the hot gas within the galaxy
(its X-ray ``atmosphere''), and is restricted to within the fiducial radius of $5r_e$.
It does not include any gas at larger radii, such as any cooler material, or any 
ICM (Intracluster Medium). We note that the ICM is most strongly present in rich clusters, but most of
our target galaxies are not in such environments.  

As hinted above, this approach to measuring both $M_X$ and $M_{tot}$ has its own set of 
intrinsic uncertainties.  For the least massive galaxies in our candidate list, the X-ray
emission falls in the low-temperature regime $k T_X \lesssim$ 0.5 keV where the Chandra 
instruments are less sensitive and the luminosity $L_X$ is also low.  
Departures from hydrostatic equilibrium in the inner regions of the bulge and halo,
due particularly to cavities and shockwaves embedded in the gas distribution,
may also be present to different degrees in different galaxies. 
Nonsphericity
of the gas profile is not a major concern:  various recent studies showed that spherical averaging of an ellipsoidal mass profile typically introduces only small biases for global quantities such as total mass and gas fraction, in X-ray hydrostatic equilibrium studies of galaxy and cluster masses (e.g., \citet{Buote12a, Buote12b} and references therein). In the case of the massive elliptical NGC 6482, 
\citet{Buote19} also found such small biases to be negligible compared to the statistical 
uncertainties.  \citet{Buote12a} (see also \citet{Churazov08}) showed that spherical averaging in a hydrostatic equilibrium analysis introduces zero bias in the inferred mass for any gas temperature profile. 
However, the baryon physics associated with assumed gas properties and uncertainties 
in the heating and cooling rates introduced by feedback may also be kept in mind \citep{Fabjan:11}.
Considering these observational limitations, we do not include in our list those galaxies 
with $kT < 0.4$ keV and $L_X < 0.4 \times 10^{40}$ erg s$^{-1}$; that is, the lowest-temperature
and lowest-luminosity systems for which the $\beta-$model fits and mass measurements are
the weakest.  Galaxies falling below those thresholds show noticeably increased scatter among the
various correlations between mass, temperature, and luminosity (see B18).  
We do, however, include NGC 1052 and NGC 1387 (both with $kT_x \simeq 0.45$ keV), 
for which the raw data are of unusually high quality and the profiles well determined.
After these cuts, we are left with 102 target galaxies.

\subsection{Stellar and Halo Masses}

For the purpose of the present analysis we have recalculated the total stellar mass
$M_{\star}$ and halo mass $M_h$ of each target galaxy, to update the previous values
used in H19 that were drawn in turn from \citet{hha2013} and \citet{hudson+2014}.
$M_{\star}$ is determined from the total K-band and V-band luminosities of each
galaxy along with mass-to-light ratios that are calibrated functions of 
integrated color (representing morphological type), as given by \citet{bell+2003}.
Specifically these are
\begin{eqnarray}
{\rm log} (M / L_K) = -0.356 + 0.135 (B-V)_0 \\
{\rm log} (M / L_V) = -0.778 + 1.305 (B-V)_0
\end{eqnarray}
where $(B-V)_0$ is the intrinsic (dereddened) integrated color.  The adopted Solar absolute magnitudes 
are $M_{K,\odot} = 3.32$, $M_{V,\odot} = 4.82$.
We adopt here the scaling for the stellar IMF defined by
\citet{chabrier2003} and \citet{kroupa2002}, using the
offsets given in \citet{bell+2003} that are needed to convert their nominal
``diet Salpeter'' IMF into the Chabrier/Kroupa scale.
Though empirically the two ways of defining $M_{\star}$ through either $L_K$
or $L_V$ are
very consistent with one another for ETGs that are dominated by
an old population with little or no recent star formation, double weight is given to the K-band estimate since it conventionally represents the total stellar mass better for
most galaxies.

Calculation of halo mass $M_h$ follows the prescriptions in \citet{hudson_etal2015}
(see Appendix C there).  Given $M_{\star}$, the stellar-to-halo mass ratio SHMR
is calculated with their ``Default'' set of functional parameters, extrapolated
to redshift $z = 0$.\footnote{The results here for $M_h$ are slightly different
from those in \citet{hudson+2014}.  In that 2014 paper, a preliminary
set of coefficients was used for the fit to the SHMR curve, but their final revised form is given in \citet{hudson_etal2015}.}  With this set of adopted parameters, their expression
for the SHMR ($= M_{\star}/M_h$) simplifies to 
\begin{equation}\label{eqn:eq7}
    {M_{\star} \over M_h} = 0.0454 \cdot \Bigl[ ({M_{\star} \over M_1})^{-0.43}
     + ({M_{\star} \over M_1})^{1.0} \Bigr]^{-1}
\end{equation}
where the pivot mass is $M_1 = 10^{10.76} M_{\odot}$.  This transformation is quite similar
in form to other recent expressions for the SHMR  \citep[e.g.][]{behroozi_etal2013,moster_etal2013,leauthaud_etal2012}
and the particular version does not affect any of the conclusions discussed below.
\citet{hudson_etal2015} implicitly use the Chabrier IMF \citep[see][]{velander+2014} so no
further adjustment to their mass scale is needed.

In Table \ref{tab2}, the compiled quantities for the observational sample of galaxies
are listed including the luminosities ($M_V^t$,
$M_K^t$), the predicted S\'ersic index $n$ (see below), 
the resulting masses ($M_{\star}, M_X, M_5, M_h$), and finally the
DM mass fraction {\bf $f_5 = f_{DM}(5r_e)$}.

\subsection{Predicted Mass Ratios from Simulations}\label{sim}

As noted above, our observational sample of galaxies ranges from nearly isolated
systems to giant central galaxies in rich environments, so it will be useful to compare the results with
simulated galaxies that cover a similar range.
To compare observations with theory, in this paper we concentrate particularly
on galaxies from the cosmological hydrodynamical Magneticum\footnote{www.magneticum.org} Pathfinder simulations (Dolag et al., in prep., but also \citet{hirschmann:2014}).  This
is a set of several cosmological simulation volumes (from $(2688~\mathrm{Mpc}/h)^3$ to $(48~\mathrm{Mpc}/h)^3$) with different resolutions extending from $m_\mathrm{Gas} = 2.6\times 10^9 M_\odot/h$ to $m_\mathrm{Gas} = 7.3\times 10^6 M_\odot/h$, performed with a modified Gadget-3 version using a WMAP7 $\Lambda$CDM cosmology \citep{komatsu:2011} with parameters $\sigma_8 =0.809$, $h = 0.704$, $\Omega_\Lambda = 0.728$, $\Omega_\mathrm{M} = 0.272$, $\Omega_\mathrm{B} = 0.0451$. Small changes in the cosmological parameters on the order of the difference between those used for Magneticum and those from Planck ($\sigma_8 =0.811$, \cite{planck6:2018}), have no significant impact on the dark matter fractions used in this work, as the changes are too small to noticeably change the dark matter halo concentration parameters; see Ragagnin et al. (submitted). 

Baryonic physics is included in the Magneticum simulations as subgrid physics, which are described in detail by \citet{teklu:2015}.  The underlying IMF used for Magneticum is a Chabrier IMF, though this does not influence the calculations of the stellar mass of the galaxies, as the IMF is only used to calculate stellar feedback and other quantities related to stellar evolution, but the stellar mass of a particle does not change once it is born.
Structures are identified with a modified version of SUBFIND \citep{springel:2001,dolag:2009}. 
For the comparison performed in this work, we use central halo galaxies from two simulations with different resolutions: 
\begin{enumerate}
    \item For a general population of field galaxies (the ``normals''), we select galaxies from Box4, which has a size of $(48~\mathrm{Mpc}/h)^{3}$ and mass resolutions for the dark matter, gas, and stellar particles of $m_\mathrm{DM} = 3.6\times10^{7} M_{\odot}/h$, 
$m_\mathrm{Gas} = 7.3\times10^{6} M_{\odot}/h$, and $m_\mathrm{*} \simeq 2\times10^{6} M_{\odot}/h$.  The gravitational softening length of this simulation at z = 0 is $\epsilon_\mathrm{DM} = \epsilon_\mathrm{Gas} = 1.4~\mathrm{kpc}/h$ for dark matter and gas particles, and $\epsilon_* = 0.7~\mathrm{kpc}/h$ for stellar particles. To ensure that the halfmass radii of the galaxies used in this work are well resolved, we use a lower total stellar mass limit of $M_\mathrm{*} \geq 2\times10^{10} M_{\odot}$ and select all central galaxies above this mass threshold. These galaxies have already been used in several publications, and for more details on their global and kinematic properties see \citet{teklu:2015, schulze:2018}. Regarding stellar masses, sizes, and dark matter fractions within the halfmass radius for the spheroidal galaxies in this sample of galaxies, we refer the reader to \citet{remus+2017} and A17. While this simulation has enough resolution to study the population of ``normal'' galaxies, the box size is too small to include massive galaxy clusters and thus BCGs, and includes only a few galaxy groups with their BGGs. Therefore, to study the ``BCG'' counterparts, we have to use a different simulation volume from the Magneticum pathfinder set:

\item For the simulated sample of centrals (BCG/BGG) in this work, we select galaxies from Box2, which has a size of $(640~\mathrm{Mpc}/h)^{3}$ and mass resolutions of $m_\mathrm{DM} = 6.9\times10^{8} M_{\odot}/h$, 
$m_\mathrm{Gas} = 1.4\times10^{8} M_{\odot}/h$, and $m_\mathrm{*} \simeq 3.5\times10^{7} M_{\odot}/h$. The gravitational softening lengths of this simulation at z = 0 are $\epsilon_\mathrm{DM} = \epsilon_\mathrm{Gas} = 3.75~\mathrm{kpc}/h$ and $\epsilon_* = 2~\mathrm{kpc}/h$ for dark matter, gas, and stellar particles, respectively. We select all central galaxies in halos with total masses
$M_\mathrm{tot}\geq5\times10^{13} M_{\odot}$ to ensure sufficient resolution. For more details on this specific simulation and its clusters see \citet{remus:2017b,lotz:2019}.  Further properties of the centrals in this sample are discussed by Remus \& Forbes (in preparation). While for this simulation the resolution is not high enough to study Milky Way like galaxies, the BCGs are well resolved.
\end{enumerate}

\noindent For both sets of simulated galaxies, the properties compared in this work are calculated in the same way. Halo masses $M_\mathrm{h}$ are calculated as the sum of all particle masses (dark matter, gas, and stars) within the virial radius, with only the substructures identified by Subfind subtracted from the halo.

Determining the ``real'' stellar masses of the simulated galaxies is, however, more of an issue. For the observed galaxies, the stellar mass $M_{\star}$ is derived from the observed luminosities, and the problem is to estimate the halo mass $M_h$ with a transformation such as the one in Eq.~\ref{eqn:eq7}. On the theory side the problem is essentially the opposite: the halo (virial) mass is well known from the simulations, but a way to estimate $M_{\star}$ needs to be defined since the total stellar mass within the virial radius (i.e. the ``real'' mass) is usually not what can be observed (especially for BCGs and their ICL). Therefore, criteria need to be applied to mimic the observational limitations. Since we do not \emph{a priori}
know the definitively correct approach to this, we consider four different ways, described below, to estimate the stellar mass in order to understand how the the dark matter fractions may be influenced. 

\begin{figure}
    \centering
    \includegraphics[width=.8\textwidth]{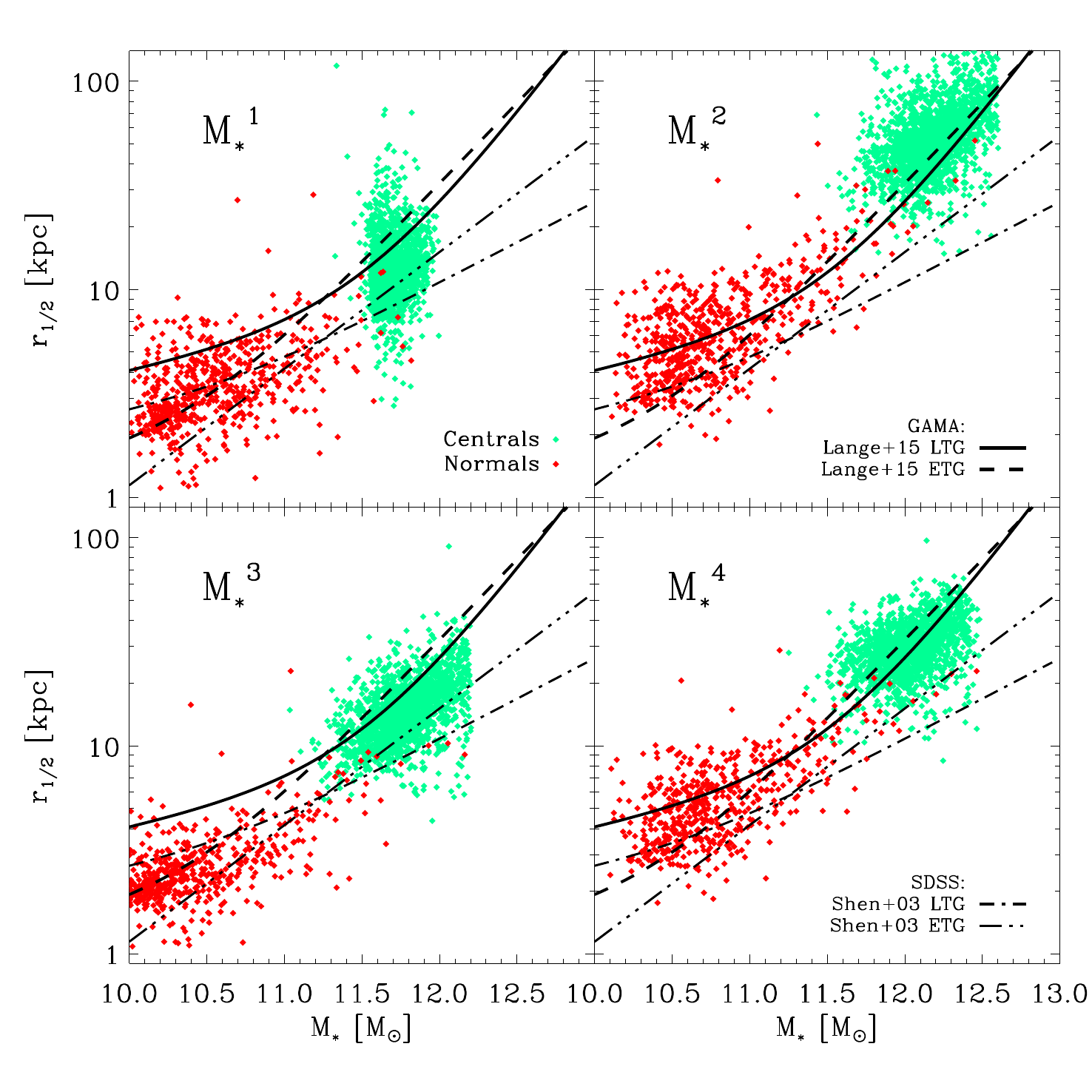}
    \caption{Stellar half-mass radius $R_\mathrm{1/2}$ versus stellar mass $M_\odot$ for the simulated galaxies, for the four different methods to calculate the stellar mass of a galaxy as described in Sec.~\ref{sim}. Red points are the ``normals'' and green points the ``centrals''. \emph{Upper left panel:}  The mass-size relation for the values obtained with the same method as used for the observations presented in this work (i.e.~by inversion of Eq.(7), method $M_*^1$). \emph{Upper right panel:}  The mass-size relation for method $M_*^2$, which is the one using all stellar particles bound to the dark matter halo.  \emph{Lower left panel:}   The results for method $M_*^3$, the one based on the split of BCGs and ICL as 40/60\%. \emph{Lower right panel:} The results for method $M_*^4$, which uses all stellar particles within 10\% of the virial radius to calculate the stellar mass.
    In all four panels, the observed mass-size relations from the SDSS survey \citep{shen:2003} for ETGs (dash-dot-dot-dotted lines) and LTGs (dash-dotted lines) and the GAMA survey \citep{lange+2015} for ETGs (dashed lines) and LTGs (solid lines) are shown for comparison.
    }
    \label{fig:sims_s_r}
\end{figure}
For each of these given stellar mass definitions we sort the stellar particles radially and sum up their masses until half of the given $M_*$ is reached. The corresponding radius defines the half-mass radius $R_\mathrm{1/2}$ as an analog to the observed half-light radius. Given the half-mass radius, the dark matter fraction within $5R_\mathrm{1/2}$ is then calculated from the particles in the simulation directly as
\begin{equation}f_5 = f_\mathrm{DM}(5R_\mathrm{1/2}) = \frac{M_\mathrm{DM}(5R_\mathrm{1/2})}{M_\mathrm{DM}(5R_\mathrm{1/2}) + M_\mathrm{*}(5R_\mathrm{1/2}) + M_\mathrm{Gas}(5R_\mathrm{1/2})}.
\end{equation} 
The four different stellar mass definition are:
\begin{itemize}
    \item \textbf{$M_*^1$:} Inverting Eq.~\ref{eqn:eq7}, we calculate the stellar mass of the galaxy from the total halo mass, analogous to the method used for the observations to calculate the total halo mass. While this approach ensures self-consistency of both observations and simulations, we also ignore the scatter in the SHMR and subsequently we under- or over-estimate the stellar mass for a significant fraction of our galaxies \citep[see][for the SHMR in Magneticum]{teklu:2017}.
    \item \textbf{$M_*^2$:} The stellar mass is calculated as all stellar particles within the virial radius with only the substructures subtracted as identified by Subfind. This method is most realistic for the field galaxies.  However, for the BCGs it adds the full mass of the ICL within the virial radius to the BCG stellar mass and therefore systematically overestimates $M_{\star}$.
    \item \textbf{$M_*^3$:} The stellar mass is calculated as 40\% of the stellar mass within the virial radius, following the average stellar mass split of 40/60 between BCG and ICL as found by \citet{remus:2017b}. This split between the BCG and the ICL is based on a decomposition of the stellar component into two populations according to their velocity distribution, with the ICL component having significantly larger velocities than the BCG. This is a good approximation for the BCGs, but is a poor approximation to the field galaxies where the stellar halos are far less than 60\% of the full stellar body of a galaxy: For example, \citet{merritt:2016} report for disk galaxies with  $M_*<1\times10^{11} M_{\odot}$ from Dragonfly an average stellar halo fraction below 1\%, and their highest fractions are still well below 10\%. Similarly, \citet{harmsen:2017} find stellar halo fractions from the GHOSTS survey of only 2-14\% of the total stellar mass.
    \item \textbf{$M_*^4$:} Following a common approach from simulations \citep[e.g.][]{teklu:2015,remus+2017,schulze:2018}, we assume the galaxy's stellar body to reside well within 10\% of the virial radius, and as such we calculate the stellar mass from all stars within $0.1 R_\mathrm{vir}$.  
\end{itemize}

Figure \ref{fig:sims_s_r} shows the resulting mass-size relations for the four different methods of deriving the stellar mass $M_{\star}$.  In this figure, we compare the four methods directly with two recent observational mass-size relations built on large samples, from the SDSS \citep{shen:2003} and GAMA \citep{lange+2015} surveys. All four definitions illustrated in Fig.~\ref{fig:sims_s_r} provide mass-size relations that are in overall agreement with the observations (though it is also worth noting that the GAMA and SDSS relations are not in close agreement with each other). Generally, all four methods used for the simulations are closer to the GAMA results than to SDSS, though method $M_*^3$ is the closest to the SDSS values, and methods $M_*^2$ and $M_*^4$ are the closest to the GAMA survey values.

Observationally, measuring the half-mass radius must confront the problem that 
the outer stellar component of galaxies
is often below the sensitivity limit of the observations and therefore the total stellar mass cannot be measured directly.  Profile fitting (and extrapolation to large radii) must be used instead.

As noted above, on the theoretical side the choice of a ``best'' approach
from the viewpoint of the mass-radius relation is not immediately clear.
Consider our method $M_*^1$ (predicting the stellar mass directly from
the halo mass and the SHMR relation) as an example.
%we demonstrate how the inferred relation differs in the mass-size relation and how its scatter changes when the half-mass radius is inferred for our simulated galaxies, where the predicted density profile is independent (and unchanged), when the total stellar mass is estimated from other observational proxies instead of using the true stellar mass from the simulations. 
Since the slope of the density profile changes with radius, a different (assumed) total stellar mass changes the inferred relation between half-mass radius and stellar mass.  In addition, due to the large scatter in the true SHMR as predicted by simulations \citep[e.g.,][]{teklu:2017}, the scatter in the stellar mass-size relation is also increased significantly if we use an incorrect stellar mass inferred from the given SHMR.  

Both the simulated and the observed mass-radius relations have
significant scatter, so we cannot definitively  exclude any of our four methods based on the mass-size relation alone. Generally, the scatter is largest for $M_*^1$ because this method neglects the intrinsic scatter in the SHMR as discussed above, but the scatter (seen in Fig.~\ref{fig:sims_s_r}) from methods $M_*^2$ and $M_*^4$ is nearly as large. Method $M_*^3$ shows the smallest scatter:  the normals are cut to only 40\% of their total stellar mass within the virial radius, which means that their stellar mass tends to be dominated by their bulges while the outer (disk) parts, which cause most of the scatter, are cut away.  Thus while this method works nicely for the centrals, it appears to be less appropriate for the normals.

One final obstacle for the simulations is that we need to compare the
half-\emph{mass} radius in 3D for the simulated galaxies to the half-\emph{light} radius
in projected 2D for the observed galaxies.  The latter is usually calculated in an optical waveband
but depends to some extent on mean wavelength. On the one hand, projected radii 
are usually smaller 
than the 3D radii by a factor of $\simeq 3/4$ for standard S\'ersic galaxy profiles \citep{ciotti1991}. But on the other hand, mass-estimated radii are smaller 
than light-estimated radii. As shown by \citet{genel+2018} for the IllustrisTNG project, 
for galaxies with stellar masses of $\log(M_*)>10.5$ these two effects approximately cancel out by chance. 
We find a similar behaviour for our Magneticum simulation sample. As projections of simulated data 
always require a (random) choice of orientation and additional uncertainty comes in when converting 
from mass to light, we simply use $R_{1/2}$ (the 3D halfmass radii) calculated directly from the simulations without
any further conversion, given that most of our galaxies have stellar masses above $\log(M_*)>10.5$.  
It should also be kept in mind that the uncertainty on $R_{1/2}$ is large enough already depending on the method used to calculate the stellar mass of the galaxies. Nevertheless, this point could be addressed in a future study in more detail.

\begin{figure}
    \centering
    \includegraphics[width=.8\textwidth]{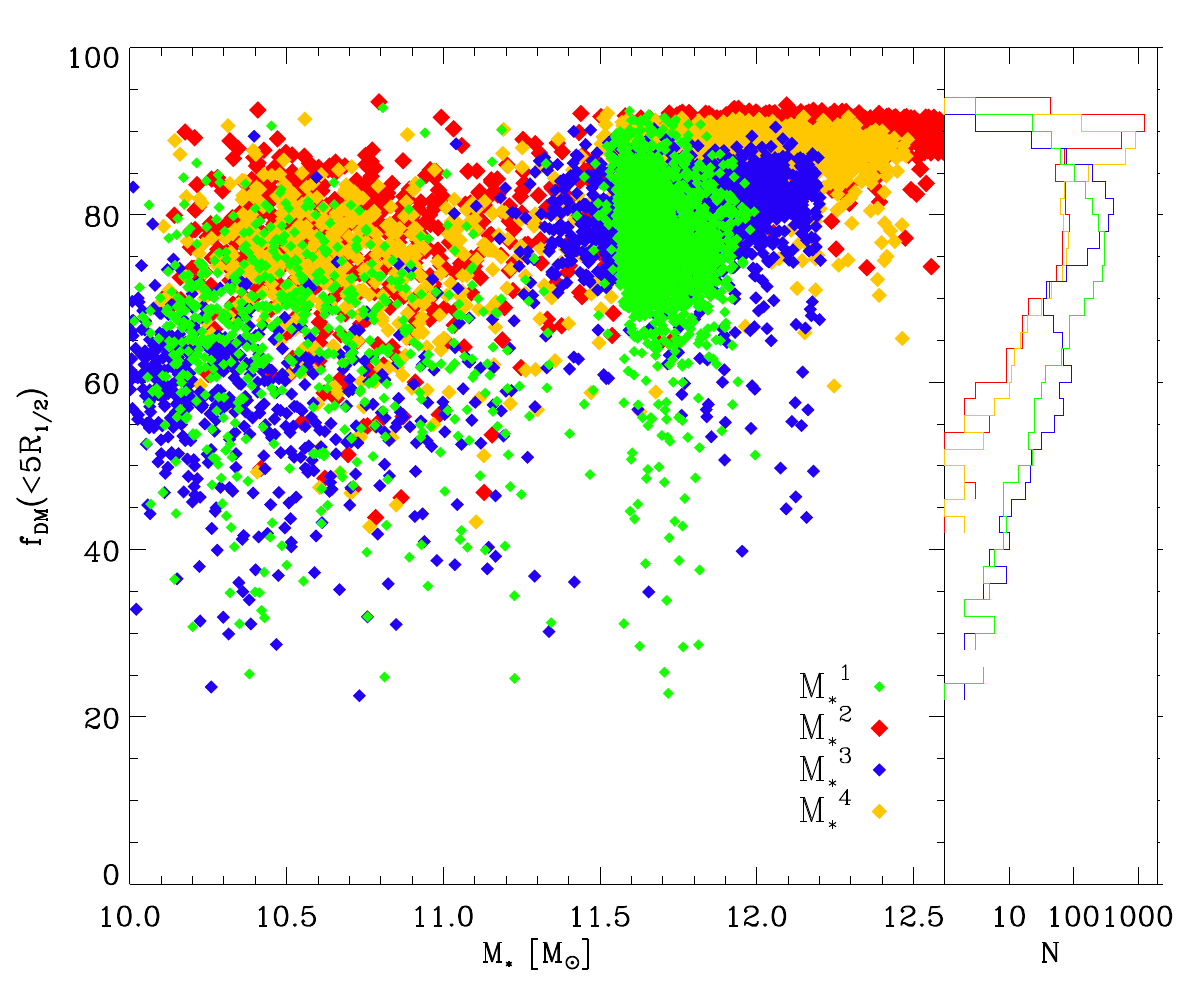}
    \caption{Dark matter fraction $f_5$ within 5$R_\mathrm{1/2}$ versus stellar mass for the simulated galaxies, with the colors indicating the different methods to calculate the stellar mass and its half-mass radius, as described in Sec.~\ref{sim}. The four different symbol colors show
    the results for the four different ways to define $M_{\star}$ defined and discussed in
    the text.  
    The right panel shows the histograms of the $f_5$ distribution for the four different methods.
    }
    \label{fig:sims}
\end{figure}

The distribution of the theoretically predicted $f_5$ values from the simulated galaxies,
for the four different ways to define $M_{\star}$, is shown in Figure \ref{fig:sims}.
Despite the very different ways in which the stellar mass is defined, the four methods show first-order agreement, with predicted $f_5$ values in the general range $\sim 0.6 - 0.9$.
Nevertheless, a small change in half-mass radius can lead both to a difference in the range of $f_5$, and also in the relative number of objects that scatter to lower $f_5$ values. For $M_*^2$ and $M_*^4$ the difference is mostly a tiny systematic shift towards lower $f_5$, clearly showing that the stellar content of all galaxies from BCGs to field galaxies is well within 10\% of the virial radius.
For the other two methods $M_*^1$ and $M_*^3$, the differences are stronger, with a much larger scatter in $f_5$. We especially see here that $M_*^3$ (which assumes that a galaxy only consists of 40\% of the total stellar mass inside a halo) provides significantly different results for the cluster environments and its BCGs and the field, normal galaxies, clearly highlighting the impact of the two distinct ICL and BCG components in galaxy clusters.
Method $M_*^1$, on the other hand, is the closest match to the observational method used in this work, and it is the method that leads to the largest scatter in $f_5$, especially at the low mass end, mirroring the large scatter in the SHMR as shown by \citet{teklu:2017}.

Table~\ref{tab_sim} lists the mean and the median of the $f_5$ distributions for the different methods of calculating $M_*$, for normals and centrals separately. In general, the BCGs have a somewhat larger dark matter fraction $f_5$ than the normals for all four methods. Additionally, for the BCGs we find that the mean and the median are approximately the same, while the mean generally is slightly lower than the median for the normals. We can also clearly see that the scatter is largest for both normals and centrals if $M_*^1$ is used.

This comparison of alternatives clearly demonstrates how sensitive the actual values of $f_5$ are to the method used to define the stellar mass of a galaxy, and the resulting differences in the halfmass radius, which are directly related \textbf{with each other} via the stellar density profile of a given galaxy. The relative differences in the slopes of the radial stellar and dark matter profiles ultimately determine how the dark matter fractions vary as a function of $M_*$ and $R_{1/2}$.
In Sec.~\ref{sec:discuss}, we will provide additional comments on the impact of the different stellar mass estimates on the resulting dark matter fractions and their match-up (or not) with the observations.

Considering all the arguments above, in the end we adopt the stellar mass definition $M_*^1$ for the simulated centrals to ensure that the ICL is subtracted properly and the scatter in the stellar mass -- halo mass relation is smallest at the high mass end.  For the normals, we adopt $M_*^2$ as we assume that for the field galaxies that any ICL component is negligible, so that essentially all stellar mass really belongs to the galaxy.  These will be our baselines for comparisons with the observations.

\section{Results and Comparisons}

The observational data in our study represent a wide range of galaxy morphology,
luminosity, environment, and other parameters.  However, our list of targets does not make up
a large enough sample (a total of little more than 100 galaxies)
to do a full analysis of DM fraction versus all these parameters.
For our present paper, we restrict the discussion to correlations of $f_5$ versus
galaxy mass, with only secondary and very broad-brush subdivisions by morphology
(early-type versus late-type) and environment (central giants versus all others).
Nevertheless, the simulations also contain a wide
variety of galaxy types and environments, so in a general sense they
represent a sample that is comparable to the real data.

\begin{table*}[t]
\begin{center}
\caption{Derived Luminosities and Masses for Observed Galaxies\label{tab2}} 
\begin{tabular}{lcccccccc}
\tableline\tableline\\
Name & $M_V^t$ & $M_K^t$ & $n$ & $M_{\star}$ & $M_X$ & $M_5$ & log ($M_h/M_{\odot}$) & $f_5$ \\
 & (mag) & (mag) & & $(10^{11} M_{\odot})$ & $(10^{11} M_{\odot})$ &$(10^{11} M_{\odot})$ & &
\\[2mm] \tableline\\
NGC193       & $ -21.80 \pm   0.34$ & $ -24.78 \pm   0.05$ & $  5.52$ & $   1.13 \pm   0.18$ & $   0.05 \pm   0.01$ & $   51.0 \pm    2.0$ & $ 12.81$ & $  0.976 \pm 0.003$ \\
NGC315       & $ -23.41 \pm   0.22$ & $ -26.38 \pm   0.04$ & $  8.27$ & $   4.74 \pm   0.59$ & $   1.98 \pm   0.05$ & $   38.0 \pm    2.0$ & $ 13.93$ & $  0.816 \pm 0.017$ \\
NGC326       & $ -23.73 \pm   0.13$ & $ -26.54 \pm   0.11$ & $  8.966$ & $   5.46 \pm   0.40$ & $   5.08 \pm   1.52$ & $  131.0 \pm   24.0$ & $ 14.05$ & $  0.917 \pm 0.019$ \\
NGC383       & $ -22.59 \pm   0.32$ & $ -25.89 \pm   0.04$ & $  6.73$ & $   2.63 \pm   0.38$ & $   0.23 \pm   0.01$ & $   52.0 \pm    2.0$ & $ 13.49$ & $  0.942 \pm 0.007$ \\
NGC499       & $ -22.04 \pm   0.40$ & $ -25.32 \pm   0.04$ & $  5.86$ & $   1.62 \pm   0.28$ & $  10.60 \pm   0.53$ & $   16.0 \pm    1.0$ & $ 13.10$ & $  0.230 \pm 0.061$ \\
NGC507       & $ -22.97 \pm   0.39$ & $ -26.03 \pm   0.03$ & $  7.41$ & $   3.17 \pm   0.52$ & $  10.40 \pm   0.69$ & $   52.0 \pm    1.0$ & $ 13.62$ & $  0.736 \pm 0.017$ \\
\\
\tableline
\end{tabular}
\end{center}
\begin{tablenotes}
\item {\sc Note:} Only the first few lines of the table are listed as a guide to form and content.
The complete table is given in the on-line version of the paper.
\end{tablenotes}
\end{table*}
\hfill \break

\begin{table*}[t]
\begin{center}
\caption{Mean and Median $f_5$ Values (simulations) for different methods to calculate the stellar mass\label{tab_sim}} 
\begin{tabular}{l|cccccc}
\tableline\tableline\\
        &            & normals      &               &            & BCGs         &              \\
 Method & Mean $f_5$ & Median $f_5$ & $\sigma(f_5)$ & Mean $f_5$ & Median $f_5$ & $\sigma(f_5)$
\\[2mm] \tableline\\
$M_*^1$   & 0.65       & 0.67         & 0.110         & 0.77       & 0.77         & 0.072 \\
$M_*^2$   & 0.77       & 0.78         & 0.075         & 0.90       & 0.90         & 0.014 \\
$M_*^3$   & 0.58       & 0.60         & 0.09          & 0.80       & 0.80         & 0.05  \\
$M_*^4$   & 0.74       & 0.76         & 0.075         & 0.88       & 0.88         & 0.024 \\
\\
\tableline
\end{tabular}
\end{center}
\end{table*}

\begin{figure}
    \centering
    \includegraphics[width=0.7\textwidth]{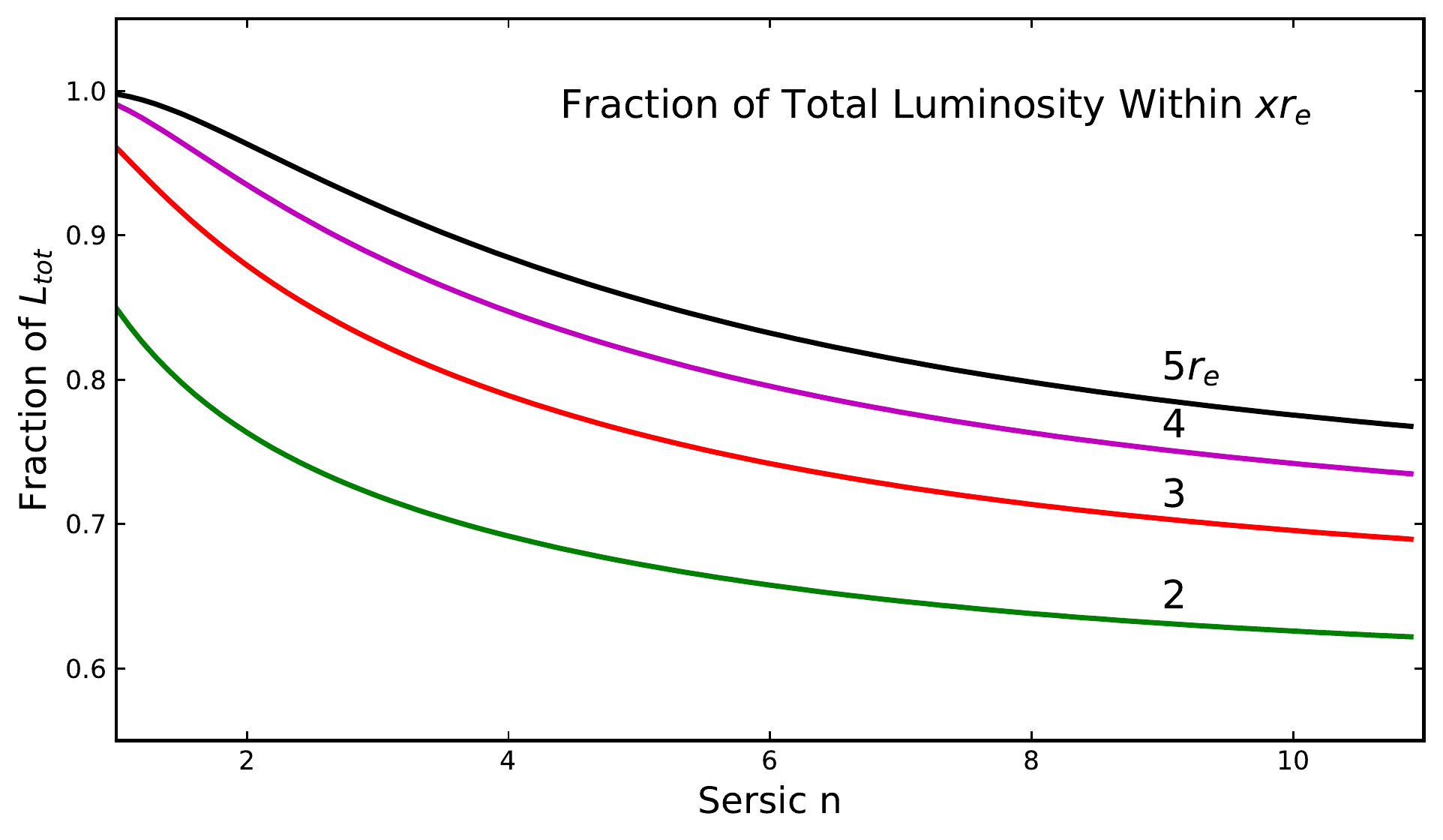}
    \caption{Fraction of total galaxy luminosity $L/L_{tot}$ contained within
    a radius $(x r_e)$, as a function of S\'ersic concentration index $n$.
    Curves are shown for $x = 2$ (green), 3 (red), 4 (magenta), and 5 (black).}
    \label{sersic}
\end{figure}

\begin{figure}
    \centering
    \includegraphics[width=1.0\textwidth]{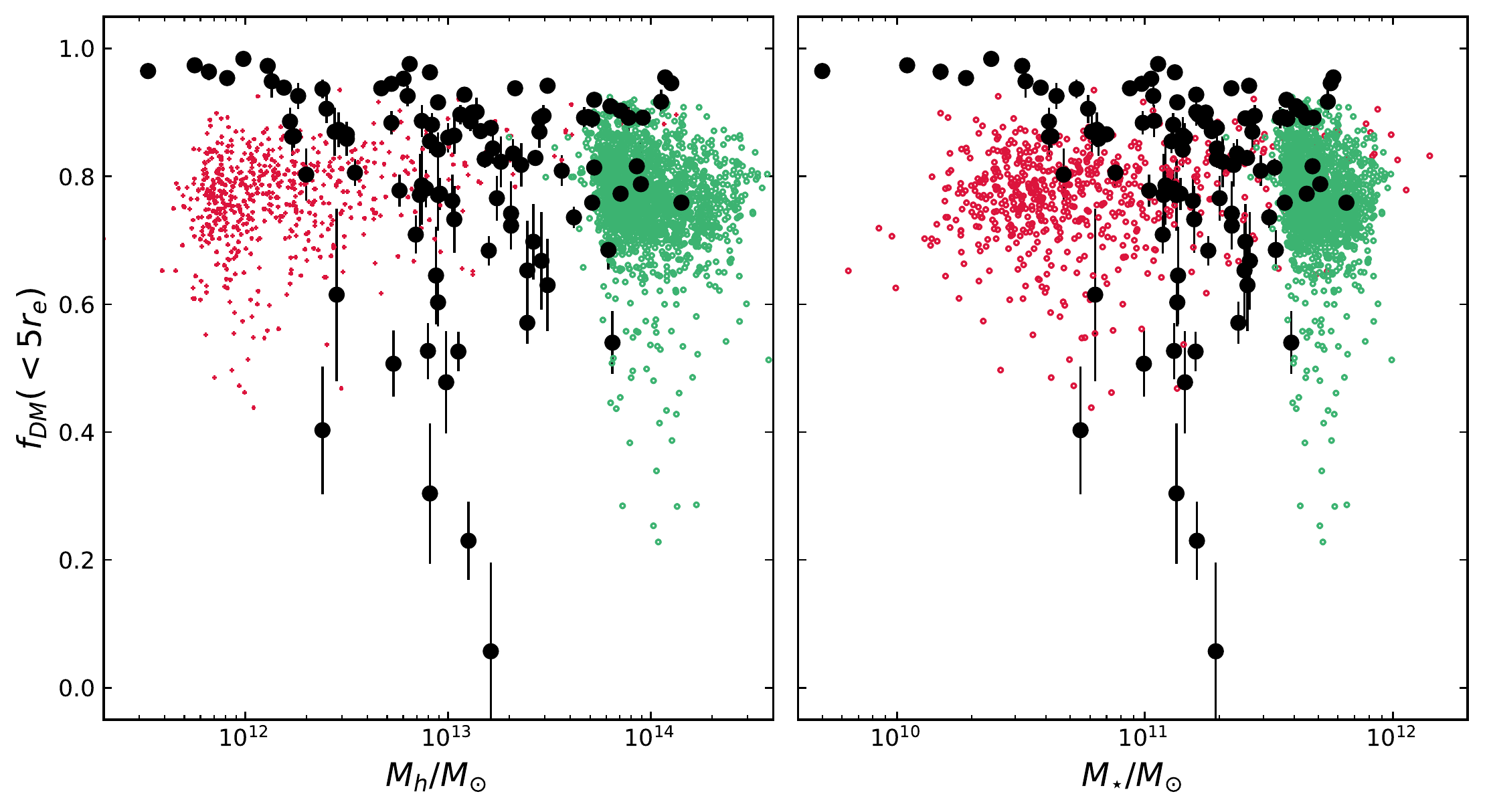}
    \caption{Dark-matter fraction within $5 r_e$ as measured from X-ray gas,
    as described in the text.  All 102 galaxies in Table 2 are shown as the
    black filled circles with errorbars.
    The DM mass fraction $f_5$ is plotted versus total
    halo (virial) mass of the galaxy (left panel) and total stellar mass
    (right panel).  
    Small dots show the results from the Magneticum/Pathfinder simulations
    \citep{remus+2017}:  the red points scattered at left are normal galaxies,
    while the green points at upper right shows the simulated centrals (BCGs/BGGs).}
    \label{figure3}
\end{figure}

\begin{figure}
    \centering
    \includegraphics[width=1.0\textwidth]{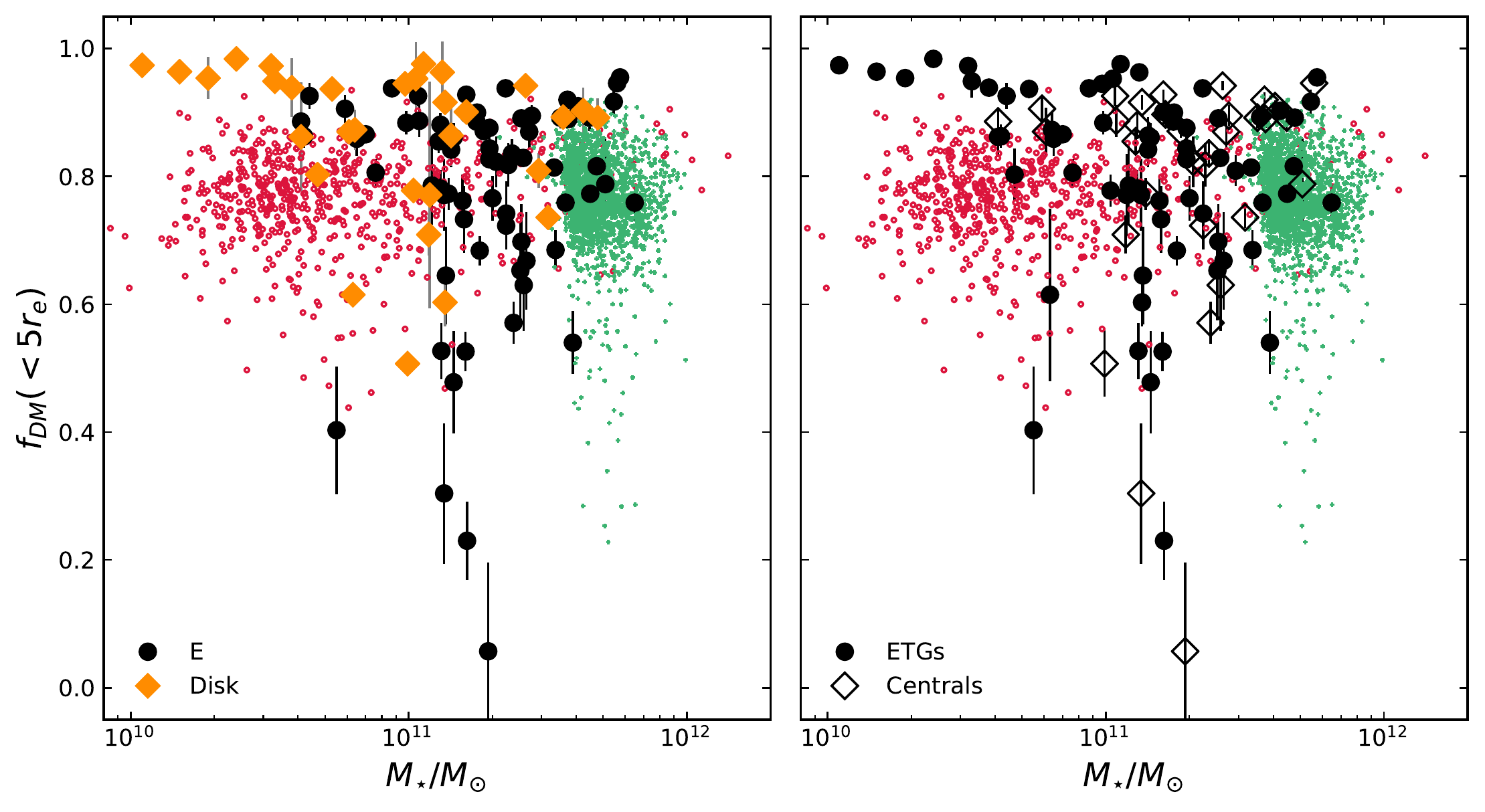}
    \caption{Dark-matter fraction within $5 r_e$ as measured from X-ray gas,
    plotted versus total stellar mass of the galaxy.  
    \emph{Left panel:} $f_5$ versus $M_{\star}$ where elliptical-type
    galaxies are plotted as black circles and disk-types in orange diamonds.
\emph{Right panel:} Central galaxies (BCGs or BGGs) are shown as open diamonds 
while other ETGs are in black circles.  
    Small dots (red, green) show the Magneticum/Pathfinder simulations
    \citep{remus+2017} as in the previous figure.}
    \label{figure4}
\end{figure}

\begin{figure}
    \centering
    \includegraphics[width=0.5\textwidth]{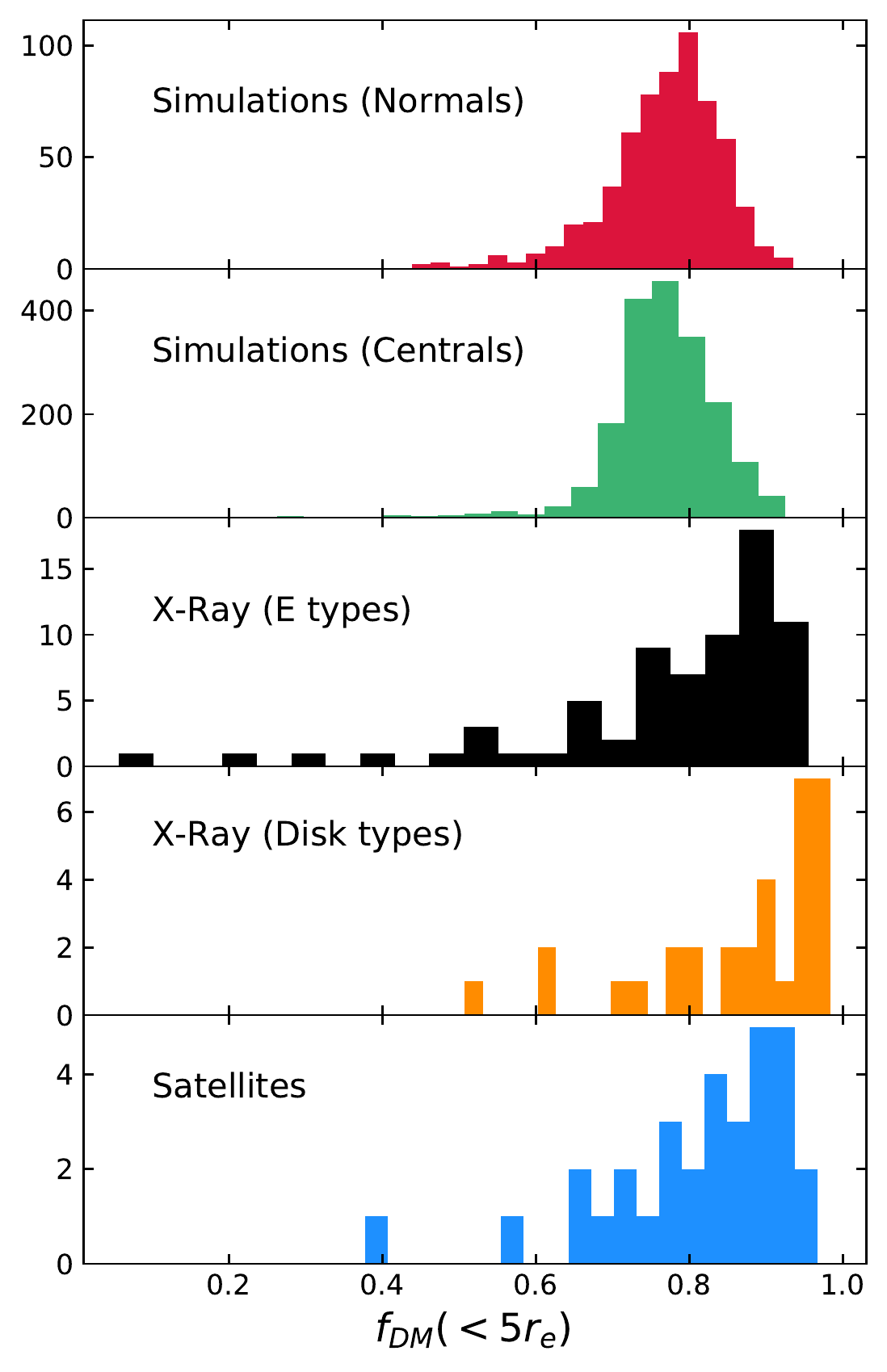}
    \caption{Histograms of mass fraction $f_5$ for the various datasets in this paper.  The Magneticum/Pathfinder simulations are shown in the top two panels, divided into
    normal galaxies and BCGs/BGGs (centrals).  
    Results from the X-ray gas method are in the two middle panels, subdivided by galaxy morphology,
    and the satellite kinematics method is in the lower panel.  
    Note particularly the numbers of galaxies with $f_5 \lesssim 0.6$ for the  
    observational methods relative to the simulations.}
    \label{histo5}
\end{figure}

Here we assume simply that $M_{bary}(r) = M_X(r) + M_{\star}(r)$, ignoring
the presence of any cool gas (which for most ETGs is small).  In most cases,
the stellar mass is in any case the dominant baryonic component.  
Lacking homogeneous detailed light profiles for most of our individual galaxies, we make the
assumption that the galaxies can be adequately described by the well known
S\'ersic profile typical for ETGs.  The central concentration
index $n$ of the profile is found empirically to increase with either $r_e$ itself
or the luminosity of the galaxy \citep[e.g.][among many others]{caon+1993,graham+1996,graham+2005,ferrarese+2006,kormendy2009,graham2019}. 
The relation between $n$ and galaxy luminosity we use here is 
(log $n = -0.104 M_V^T - 1.56$) for galaxy total
absolute magnitude $M_V^T$ \citep{kormendy2009,graham2019}; it is closely
consistent with the other studies cited above, and gives predicted $n-$values
that are entirely consistent (within the large empirical scatter)
with (e.g.) observed correlations of
$n$ with $r_e$ or stellar mass $M_{\star}$.  These calculated values of $n$
are listed in Table \ref{tab2}.

We denote $q_n$ as the fraction of the total light contained within $5 r_e$.  
In Figure \ref{sersic}, $q_n$ as calculated by integration of the S\'ersic profile 
is shown versus index $n$, for fiducial radii of $r = 2$, 3, 4, and $5 r_e$ 
(though only $5 r_e$ is relevant for the discussion to follow); 
by definition, $q_n = 0.5$ for $1 r_e$.
For the classic de Vaucouleurs profile ($n=4$), 88\% of the total light falls
within $5 r_e$, but for the largest observed $n-$values such as apply to
the very luminous, extended BCGs, $q_n$ falls to 78\% or less.
For the galaxies with
$M_V^T \lesssim -19.5$ that are in the present study, to a good approximation
we have 
\begin{equation}
    q_n = 1.056 - 0.278 ~{\rm log}~n \,.
\end{equation}

The DM mass fraction can then be defined as 
\begin{equation}
    f_5 = 1 - {(q_n M_{\star} + M_X) \over M_5 } \, .
    \label{eqn:f5}
\end{equation}
The final calculated $f_5$ values are listed in the last column of Table 2.
In practice, $f_5$ is insensitive to $n$ since $M_5$ dominates over $M_{\star}$:  even the
simple assumption of a de Vaucouleurs profile ($n=4$) for all galaxies would
change the $f_5$ estimates in Table 2 by $\pm0.02$ at most.

The next stages are to plot the
observed distribution of $f_5$, compare with the simulated galaxies,
and look for any dependencies on galaxy mass, morphological type, or environment.
In Figure \ref{figure3}, $f_5$ as calculated from the X-ray sample of 102 galaxies is plotted versus both halo and stellar mass
($M_h, M_{\star}$).  In addition, the predicted $f_5$ values for the realizations of the Magneticum Pathfinder simulations described above (both normal galaxies and centrals)
are shown for comparison.  Not surprisingly, one effect of observational selection is immediately seen:
the majority of the observed galaxies lie at fairly high masses where $T_X$ and $M_X$ are larger 
and more securely measurable, whereas the simulated `normal' galaxies follow a luminosity function that more heavily populates
the lower-mass end of the distribution regardless of how much gas they contain.  
For this reason, only the trends of $f_5$ versus mass or type
are relevant, and not the relative numbers of galaxies in a given mass range.

In general, good first-order agreement is seen between the mean level
of the simulations and the observed galaxies, as well as the typical scatter.
The observed galaxies, however, show excursions to both higher and lower
$f_5$ levels ($\gtrsim 0.9$, $\lesssim 0.6$) that are very rarely reached
in the simulated systems.  Both of these extremes will be discussed below.
Notably, no strong trend of $f_5$ versus galaxy \emph{mass} is evident,
at least in the mass range studied here.

Though both $M_h$ and $M_{\star}$ are shown in Fig.~\ref{figure3}, they give
much the same information about the distributions of DM fraction versus galaxy mass.  In what follows, 
we will therefore show only $f_5$ versus stellar mass $M_{\star}$, though if desired, any of the distributions
against $M_h$ can be constructed from the information in Table 2.

In Figure \ref{figure4} (left panel), the galaxies are separated into elliptical (E) and disk
(S0/SA0/SB0) types to see if any obvious differences emerge versus
morphology. 
A particularly noticeable subset consists of the $\sim 10$ disk
galaxies at upper left in Fig.~\ref{figure4} that have $f_5 \simeq 0.95$;
these are also the lowest-mass ones in the sample ($M_{\star} < 5 \times 10^{10} M_{\odot}$).
A17 found a slight trend in the opposite sense between S0 and E types, 
though as they point out, any
trends with galaxy type or environment in their sample were obscured by the large
dispersion in $f_5$ and small-sample statistics.
These very DM-dominated disk galaxies are reminiscent of the
disk systems studied in the GHOSTS and Dragonfly surveys, some of which have very sparse stellar halos \citep{merritt:2016,harmsen:2017}.  We note as well that even within
$1 r_e$ the halos of some moderately luminous ETGs are quite DM-dominated
\citep{tortora+2019} and thus should be even more so at $5 r_e$.  Interpretations from galaxy formation simulations
\citep[e.g.][]{elias+2018,d'souza+2018,monachesi+2019} suggest that systems in this
high$-f_5$ range are likely to have had quieter evolutionary histories, with earlier and higher fractions of \emph{in situ} star formation.  The Magneticum
simulations for the `normal' suite of galaxies, however, predict very 
few galaxies with $f_5 > 0.9$.

Another noticeable subset consists of the 
four galaxies with $f_5 \lesssim 0.4$:  these are NGC 499, 4697, 5044, and IC 1262.  In all cases these are ones where the estimated gas mass $M_X$ is extremely high,
and perhaps implausibly so.  While these four objects are kept in the lists
for the present, we suggest that these cases call for remeasurement of
the gas mas with higher signal-to-noise observations. 
It is worth noting as well that the effect of errors in $M_X$ is strongly asymmetric.
If $M_X$ is severely {\it under}estimated, $f_5$ would increase by only a tiny 
amount because $M_{\star}$ already dominates over $M_X$ in most cases.  
But if $M_X$ is {\it over}estimated, $f_5$ can decrease very significantly,
as we see here for these few cases.

In this context, we note that a few of the $f_5$ 
values measured through satellite dynamics (A16, A17) 
were also found to lie at similarly low
values in their papers, but once their adopted IMF scale is renormalized (see below), these move up into
the normal range $f_5 \gtrsim 0.5$ occupied by the majority of cases (see the 
discussion in the next section).

In Figure \ref{figure4} (right panel), the same data are displayed but 
now broken out roughly by environment:  BCGs or BGGs
are contrasted with all others that are not the centrally dominant objects in their
local environments.  The BCGs/BGGs almost all have masses $M_{\star} > 10^{11} M_{\odot}$.  No strong difference in the $f_5$ distributions is evident between them and the normals.
Apparently, central location by itself does not correlate with unusually high or
low DM fraction, at least in the same mass ranges.  Notably, the simulated centrals lie in very much the same $f_5$
range as their real-world counterparts, at $f_5 \simeq 0.7-0.9$.  The same result
was found by A17 from the 7 BCGs in their list.

\begin{figure}
    \centering
    \includegraphics[width=0.8\textwidth]{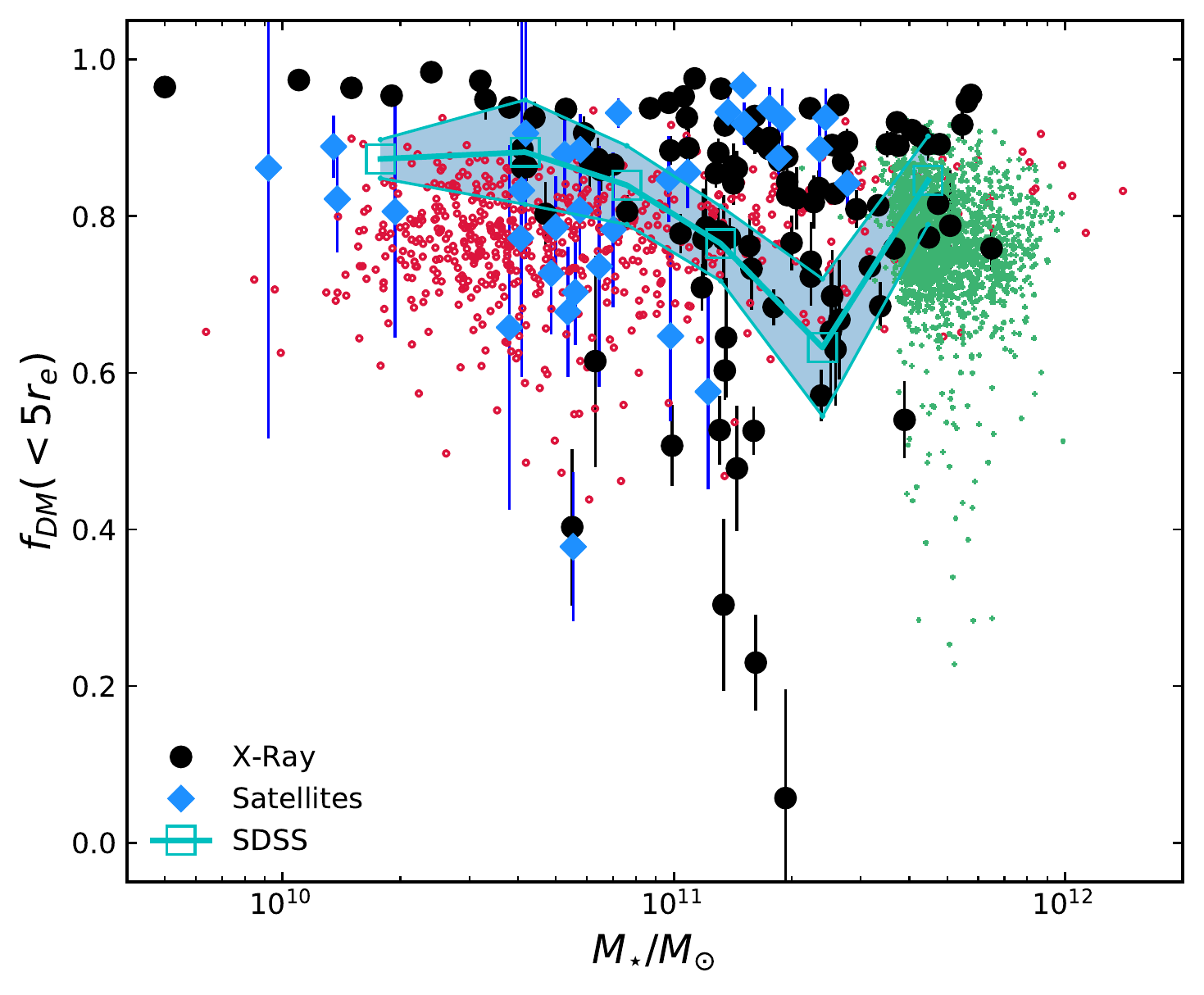}
    \caption{Dark-matter mass fraction $f_5$ for our entire sample of
    X-ray galaxies (black filled circles) is compared with the results from analysis
    of satellite dynamics \citep[blue diamonds, from][]{alabi_etal2017}.  The large open squares
    and the cyan-shaded region show the mean results from the SDSS galaxy sample studied
    by \citet{wojtak_mamon2013}.}
    \label{compare}
\end{figure}

\begin{figure}
    \centering
    \includegraphics[width=0.6\textwidth]{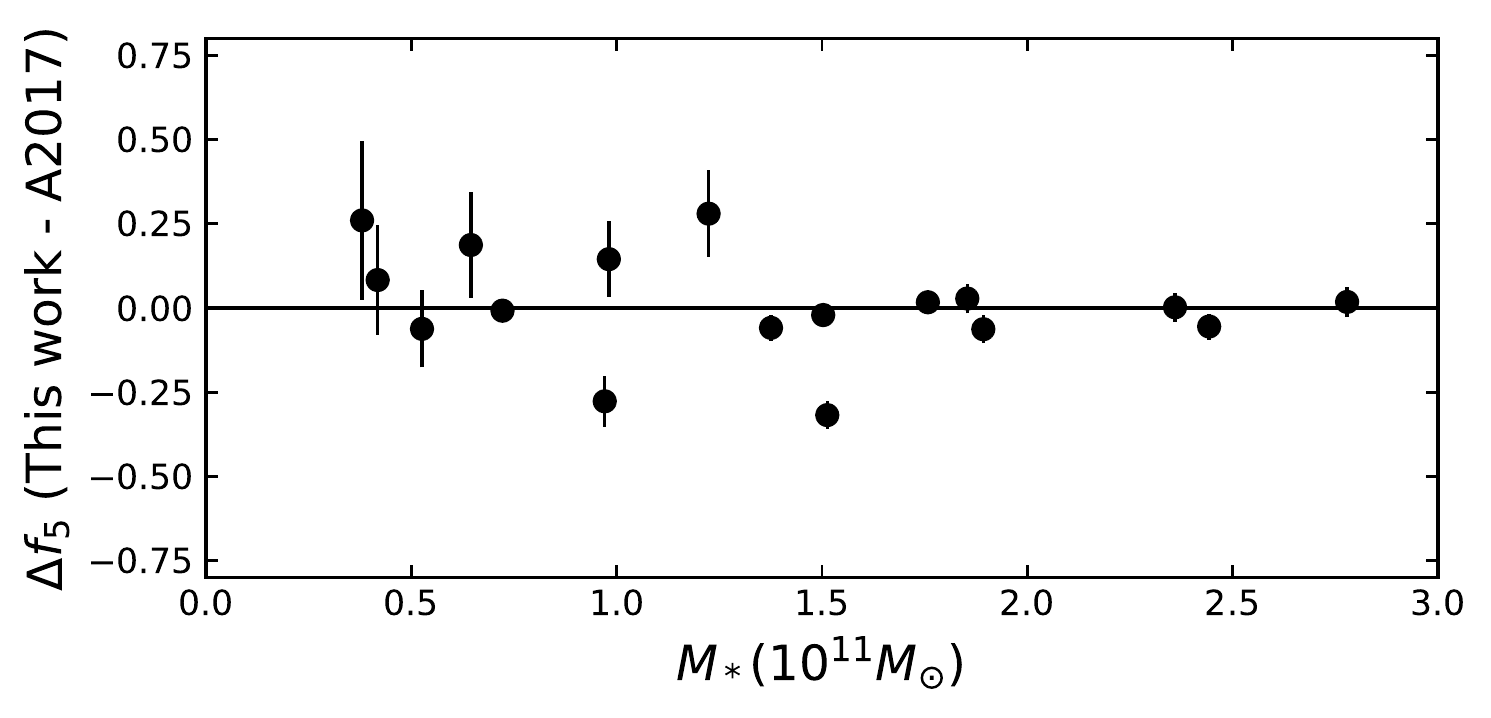}
    \caption{Comparison of our results for $f_5$ with those from A17, for
    the overlapping set of 17 galaxies.  The difference $\Delta f_5$ (This work
    minus A17) is plotted versus stellar mass $M_*$.
    }
    \label{delta-f}
\end{figure}

The distributions of $f_5$ are shown
in histogram form in Figure \ref{histo5}.  Separate panels are used to
show the two groups of simulated systems (normal or central), and two groups
of the X-ray observed galaxies (E-types and disk types).
The difference between the E-type and disk-type systems is more evident here: 
the disk systems sit higher on average, and they lack the `tail' extending to lower $f_5$. 
The $f_5$ distributions for the E and Disk samples 
turns out to be statistically significant at $98$\% confidence 
according to a KS test.

\begin{table*}[t!]
\centering
\begin{threeparttable}
\caption{Mean and Median $f_5$ Values (Observations)\label{tab3}}
\centering
\begin{tabular}{lcccc}
\hline
 && \\
 Sample & $N$ & Median $\overline f_5$ & Mean $\langle f_5 \rangle$ & $\sigma(f_5)$ \\
\\
\hline
&&\\
X-ray, E-type & 70 & 0.83 (0.02) & 0.78 (0.02) & 0.17 \\
X-ray, Disk-type & 32 & 0.90 (0.03) & 0.87 (0.02) & 0.12 \\
X-ray, BCG/BGG & 33 & 0.87 (0.04) & 0.80 (0.03) & 0.20 \\
Satellite Dynamics & 32 & 0.85 (0.03) & 0.81 (0.02) & 0.12 \\
X-ray + Satellites & 17 & 0.89 (0.04) & 0.85 (0.03) & 0.12 \\
Combined Best & 117 & 0.86 (0.02) & 0.81 (0.02) & 0.15 \\
%Simulations (Normal) & 622 & 0.67 (0.01) & 0.65 (0.01) & 0.115 \\
%Simulations (BCG) & 1914 & 0.77 (0.004) & 0.77 (0.003) & 0.072 \\

\\
\hline
\end{tabular}
\end{threeparttable}
\end{table*}

\section{Discussion}\label{sec:discuss}

In Table \ref{tab3}, we list some numbers that roughly characterize each 
sample:  the sample median $\overline f_5$, and the sample
mean $\langle f_5 \rangle$ and rms dispersion
$\sigma(f_5)$ along with their uncertainties.
Because the $f_5$ distributions are asymmetric, the median is higher than the
mean, though in no case by more than $\simeq 0.1$.  

As noted above, the simulated galaxies occupy
a range of $f_5$ that to first order  
matches the observed samples, though the match is best for the giant centrals.  
Given both the present state of development of the simulations and the 
measurement uncertainties for the real galaxies, it is not yet clear that systematic
differences at the level of $\Delta f_5 \lesssim 0.1$ between the observations
and the simulations in any part of the $(f_5, M)$ plane can be viewed as  
significant.  It seems quite possible that differences of this order can be
the results of the basic differences in the way that $M_5$ in particular is 
calculated, but could also originate from our choice of calculating the stellar masses from simulations, as discussed previously and as also seen from Table \ref{tab_sim}. 
Further discussion on these points will be made below.

\subsection{Comparison with Satellite Dynamics Methods}

In Figure \ref{compare}, the $f_5$ distribution from the X-ray gas method is now compared
more completely with estimates from satellite dynamics (GCs and PNe) for 32 galaxies, as listed by A17.  
Their adopted distances employ the same distance scale \citep{brodie+2014} and agree closely with
ours.  However, their analysis assumed a mass-to-light ratio $M/L_K = 1.0$ that is
roughly equivalent to a Salpeter IMF.  Their adopted $M_{\star}$ values have therefore been
adjusted by $-0.3$ dex to put them close to the Chabrier IMF that we use here.
After this renormalization, the results are as shown in Figure \ref{compare} for the 102 X-ray measurements
(black dots) and the 32 satellite-dynamics measurements (blue diamonds).
In a strict sense, the A17 datapoints are upper limits to $f_5$ since they include
only stellar mass and not gas mass in $M_{bary}(5 r_e)$; however, in most cases
the difference is small since $M_{\star}$ dominates (see H19).  
A KS test shows that the satellite-$f_5$ distribution is not significantly
different from the X-ray sample.

The satellite-based data make an especially important contribution to filling in the 
mass range $M_{\star} \lesssim 10^{11} M_{\odot}$ where relatively 
few galaxies contain X-ray gas
that is easily measurable. 
The A17 set of galaxies does not constitute
an entirely independent list from the X-ray targets, however.  There are 17 galaxies
in common between the lists, for which H19 compared the $f_5$ and $M_5$ values
(see their Figure 6).\footnote{H19 actually compared 20 galaxies measured by both
methods.  However, in the present paper, a few of these were removed from our sample because of their low $L_X$ and $T_X$ as discussed above.} 
In Figure \ref{delta-f}, for these 17 overlapping galaxies the difference 
$\Delta f_5$(Xray-A17) is plotted versus stellar mass. To make the two datasets
more strictly comparable, the gas mass $M_X$ has been removed from the X-ray measurements
of $f_5$ before calculating $\Delta f$..  At the highest masses the
agreement between the two methods is close with little scatter; at the lowest
masses the scatter increases, but the net offset $\Delta f_5$ is still consistent with zero
and is independent of $M_{\star}$.
In brief, we find no serious evidence of systematic disagreement between the
two methods, once both have been put onto the same IMF scale.
Interestingly, the median of the 17 overlapping galaxies 
is $\overline f_5 = 0.89 \pm 0.04$ with no low$-f$ outliers.

Lastly, the results from \citet{wojtak_mamon2013} are added for comparison with both the
X-ray and A17 data.  Wojtak \& Mamon used spectroscopy and photometry from the 
Sloan Digital Sky Survey (SDSS) 
to extract 3800 isolated nearby ETGs along with their satellite \emph{galaxies} to deduce 
the mass profiles along with $f_1, f_2,$ and $f_5$.  These systems were selected such that
their satellites were $\gtrsim 1.5$ magnitudes fainter than the central galaxy, and
so they can be seen as fitting into
the BGG category or even as ``fossil groups''.  Their method uses an anisotropic model for the
phase-space distribution of satellites, generalized for the case where the DM halo and the
tracers may follow different radial profiles.  Their mean points for $f_5$ are tabulated
in 6 bins of galaxy stellar mass and are shown in Fig.~\ref{compare} as the shaded region.
They use the Chabrier IMF scale and so their data need no adjustment before comparison.
Just as for A17, however, their $f_5$ estimates do not include gas mass in $M_{bary}$ and so in 
a strict sense are (slight) upper estimates.  A striking feature of the SDSS sample 
is the clear dip in the mean $f_5$ near $M_* \sim 3 \times 10^{11} M_{\odot}$, very near
the mass range where we see most of the low$-f_5$ outliers
in the X-ray and satellite methods.

\begin{figure}
    \centering
    \includegraphics[width=0.8\textwidth]{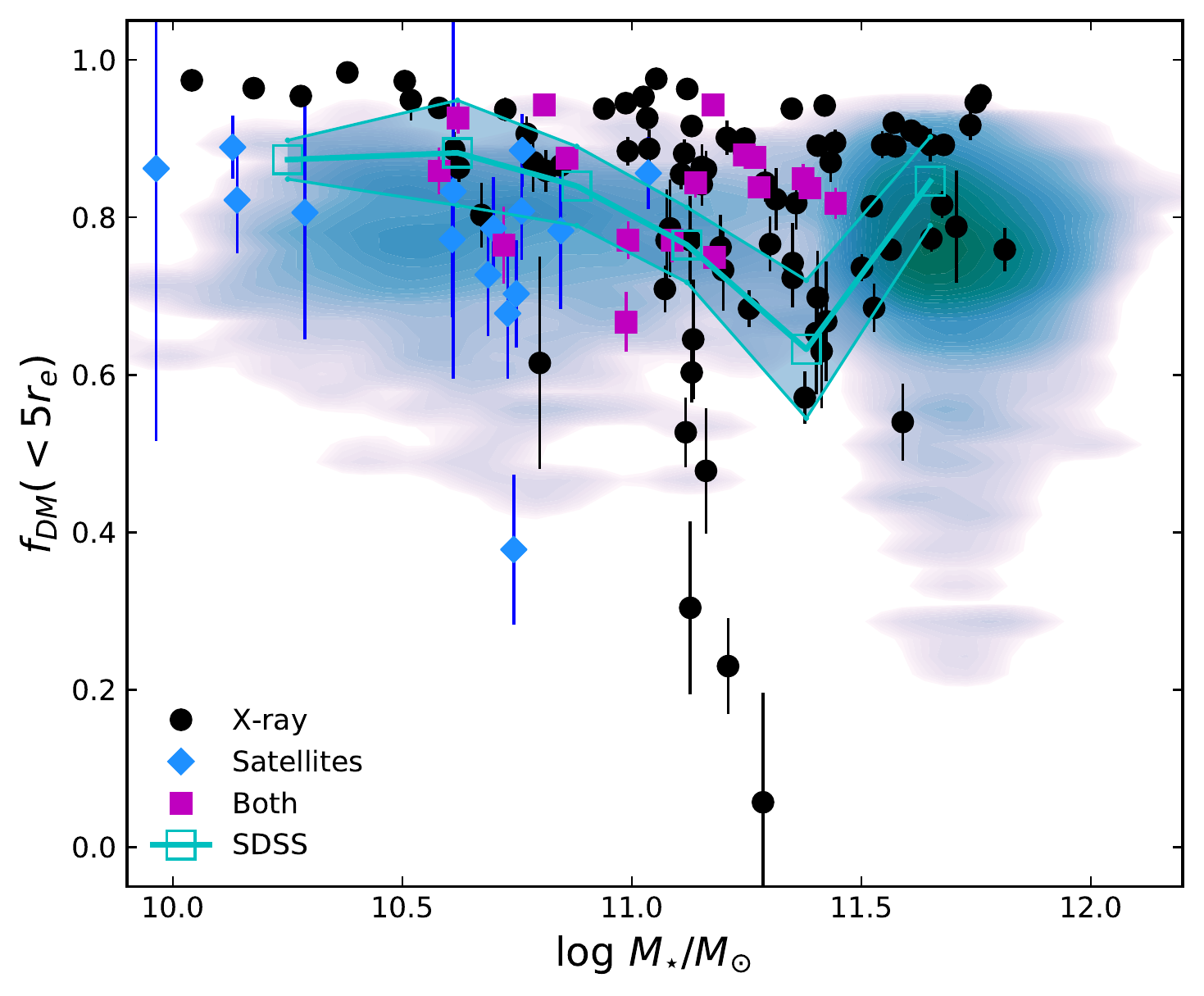}
    \caption{Dark-matter mass fraction $f_5$ for all measurements combined.
    Black filled circles represent 85 systems measured through the X-ray
    method alone; blue diamonds the 15 systems through satellite
    dynamics alone; and magenta squares the 17 systems for which both
    methods are averaged (see text).  The shaded region with mean lines shows the results
    from the SDSS satellite analysis as in the previous figure.  The filled-contour plots show the
    distribution of the simulated systems; contour levels have been logarithmically
    scaled for visibility.}
    \label{final}
\end{figure}

In Figure \ref{final} the distribution of $f_5$ is shown for a ``best'' dataset 
constructed from combination of the X-ray and satellite-dynamics methods.  For
the 17 galaxies in common between the A17 list (n=32) and the X-ray list
(n=102) we take an average of the measurements, since there
is no compelling physical reason to prefer one method over the other.
In combination the total is 117 galaxies:  17 measured by both methods, 15 from satellite
dynamics only, and 85 from X-ray only.

\begin{figure}
    \centering
    \includegraphics[width=0.6\textwidth]{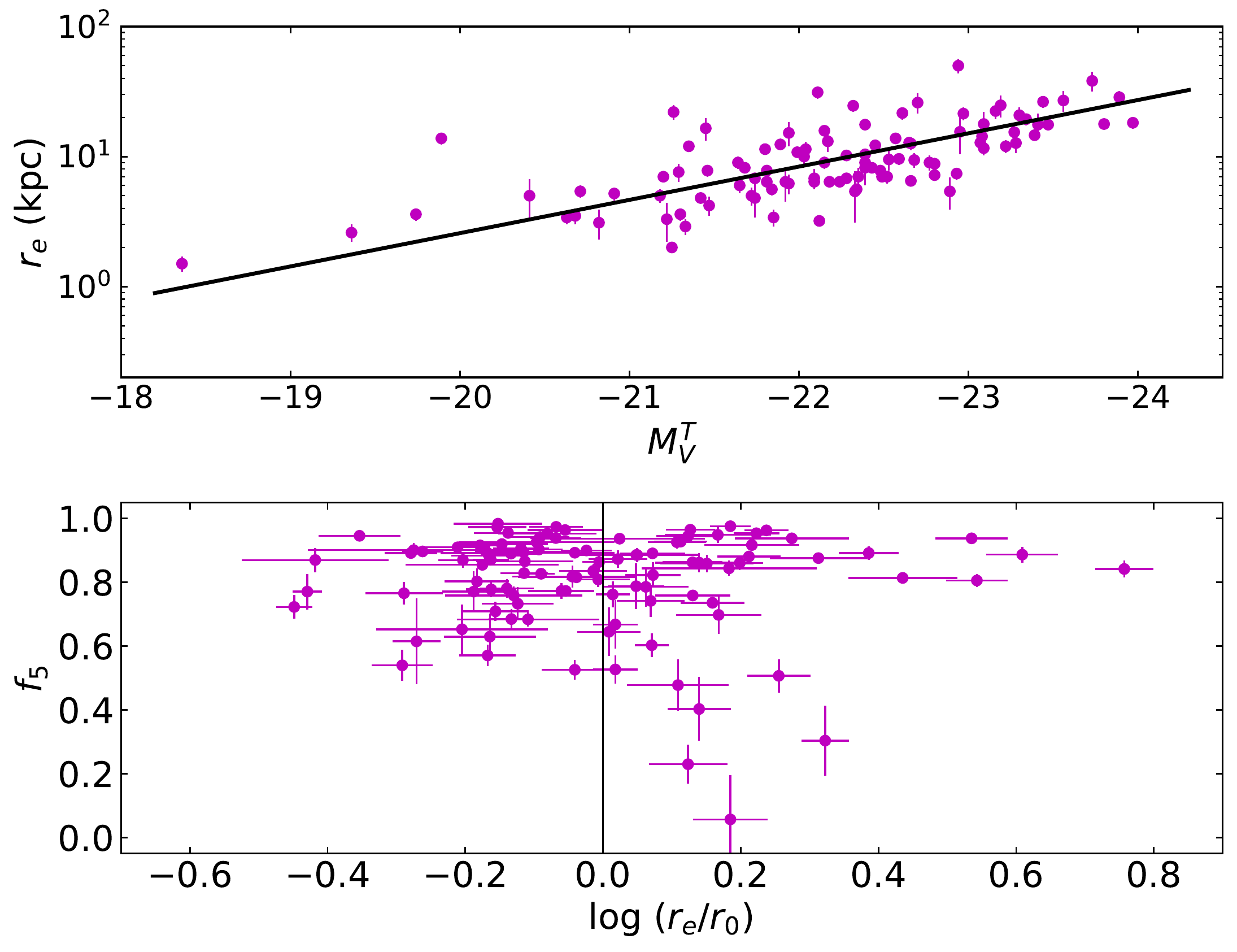}
    \caption{\emph{Upper panel:} Effective radius $r_e$ (kpc) versus 
    luminosity $M_V^t$.  The \emph{solid line} is not a fit to
    the datapoints:  it shows the normal relation for ETGs from 
    \cite{bender+2015}, as given in the
    text.  \emph{Lower panel:} Scatter plot for $f_5$ (from Table 2)
    versus normalized effective radius $(r_e/r_0)$, where $r_0$ is
    the expected effective radius for each galaxy's luminosity from
    the normal relation (see text).
    The vertical line denotes $r_e = r_0$.}
    \label{renorm}
\end{figure}

%\begin{figure}
%    \centering
%    \includegraphics[width=0.6\textwidth]{re_mass.pdf}
%    \caption{\emph{Upper panel:} Effective radius $r_e$ (kpc) versus 
%    stellar mass $M_{\star}$ for the 102 in our X-ray sample.  
%    Two empirical relations from previous literature are shown
%    for comparison, including \citet{lange+2015} (solid line)
%    and \citet{mosleh+2013} (dashed line).
%    \emph{Lower panel:} Our measured effective radius $r_e$ versus
%    Sersic index $n$ predicted from $M_V^T$ as given above.  A standard
%    relation from \citet{caon+1993} is shown as the solid line.}
%    \label{remass}
%\end{figure}

\begin{figure}
    \centering
    \includegraphics[width=0.8\textwidth]{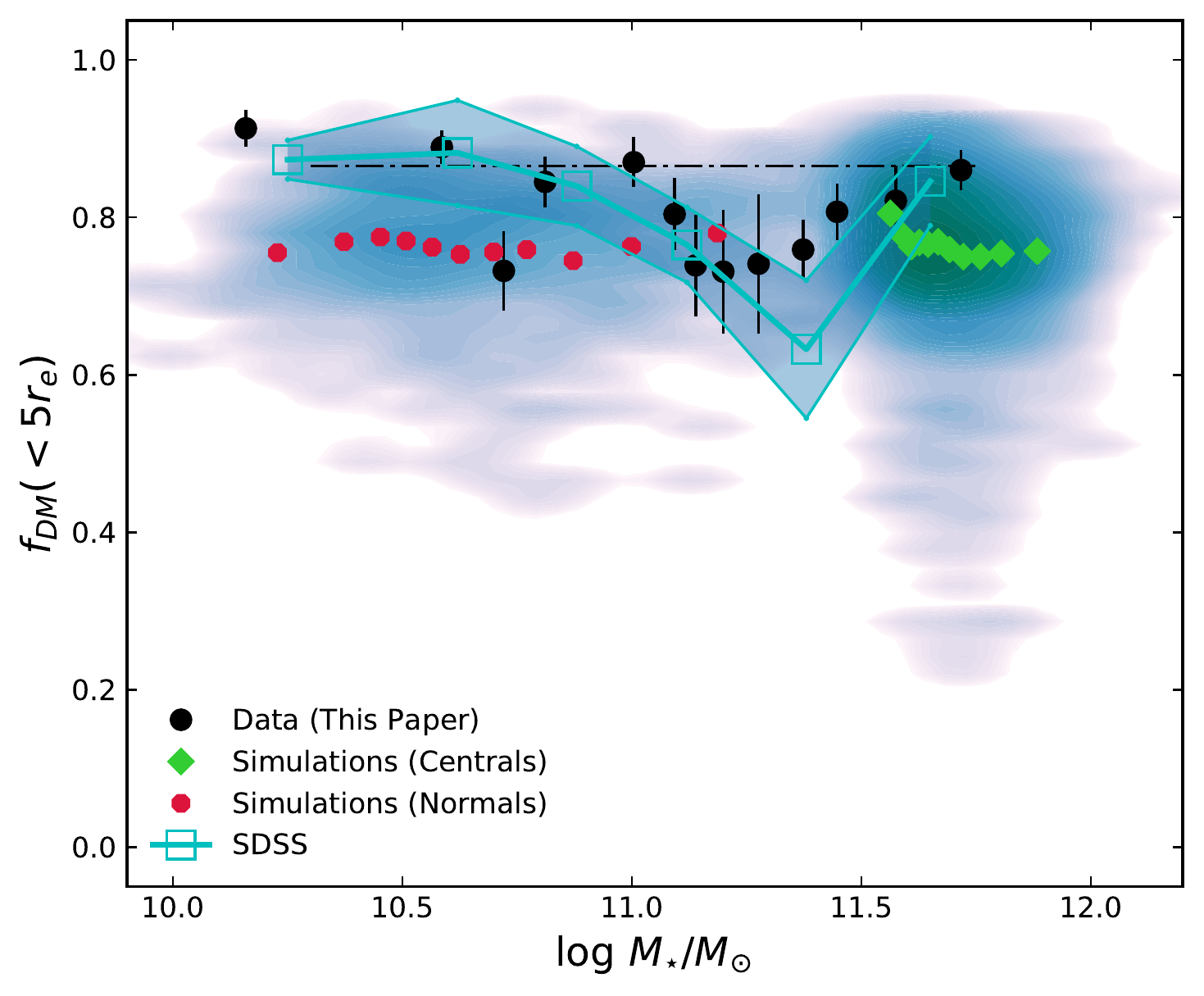}
    \caption{Mean datapoints for $f_5$, grouped in bins of $M_{\star}$.
    Black filled circles with errorbars show the mean points from the total sample
    of 117 galaxies combining both X-ray and satellite data (see text).
    The shaded cyan region with mean lines shows the results
    from the SDSS satellite analysis as in the previous figures. The filled green 
    diamonds show the mean points for the simulated central (BCG/BGG) galaxies, with stellar masses
    $M^1_{\star}$ defined by Eq. (7).  The filled red circles are the
    mean points for the simulated galaxies with stellar masses defined by $M^2_{\star}$ (including all stars) as described above.  Finally, the
    dot-dashed horizontal line at $f_5 = 0.86$ is the mean value obtained through 
    the strong-lensing technique from a sample of 161 ETGs \citep{oguri+2014}.}
    \label{meanpoints}
\end{figure}

\begin{figure}
    \centering
    \includegraphics[width=0.8\textwidth]{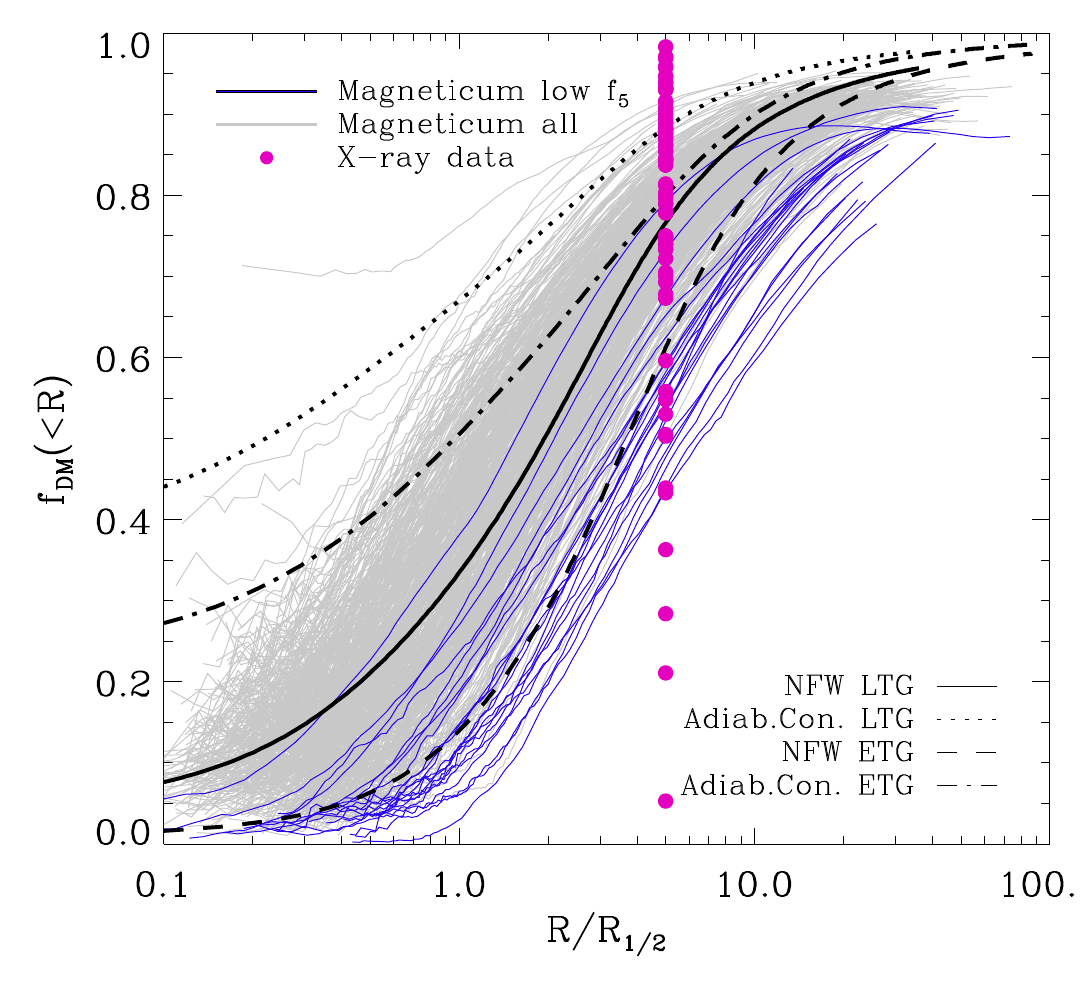}
    \caption{Radial dark matter fraction profiles for all normal galaxies from the Magneticum simulations, with the halfmass radius $r_{1/2}$ calculated using method $M_*^2$, e.g. all stars inside the virial radius. Those galaxies with dark matter fractions below $f_5<0.6$ in this method, that is below $f_5<0.6$ using method $M_*^2$, are shown as blue lines. The observations from the X-ray method presented in this work are shown as magenta dots.
    Black lines mark the calculated radial dark matter fractions from \citet{courteau+2015} assuming a NFW profile around a LTG (solid line), a NFW profile around an ETG (dashed line), a contracted dark halo around a LTG (dotted line), and a contracted dark halo around an ETG (dash-dotted line).}
    \label{expanded}
\end{figure}

\subsection{Scatter and the Effects of Measurement Uncertainties}

On strictly observational grounds, unphysical scatter in the estimated $f_5$ values can be generated
in a variety of ways, but perhaps most importantly from measurement uncertainty in $M_5$,
the total gravitating mass within $5 r_e$.
In the defining relation for $f_5$ (Eq.~\ref{eqn:f5}), $M_{\star}$ is fairly well understood 
except for some BCGs with extended-halo
envelopes that may require multiple radial components for a fit.  
$M_X$ may be internally uncertain, but is also usually small compared with $M_{\star}$.
Thus in most cases, the calculated uncertainty in $f_5$ is dominated
by the uncertainty in $M_5$.
Any under/overestimate of $M_5$ translates directly into under/overestimation of $f_5$.
In turn, $M_5$ is sensitive to the measured values of $\beta, T_X$, and $r_e$,
varying in nearly direct proportion to all three.

The estimated values of $\beta$ and $T_X$, as mentioned above, depend on the properties 
of the fits to the X-ray data:  the assumptions are that the gas is 
isothermal, it is in hydrostatic equilibrium, a single $\beta-$model profile is valid, and
spherical symmetry applies (Eqns.~(2-4) above).  These assumptions differ in degree 
of validity from case to case (see B18).  Similarly, the satellite-dynamics method for
deriving $M_5$ has its own, very different, uncertainties including the degree of substructure
and correlated motions in the tracer particles, the orbital anisotropy profile, and 
small-number statistics (see A16, A17).  

\subsection{Effective Radii}

One factor that is common to both X-ray and satellite methods is the uncertainty
in the key quantity $r_e$, which affects all of $M_{\star}, M_X$, and $M_5$.  
Though very simple in principle (the projected radius enclosing half
the total stellar light), $r_e$ is difficult to measure consistently and different 
methods have a long history of
internal disagreement and scatter.  Graphic examples of such scatter are shown in \citet{gharris_etal2014}
(see their Fig.~2) where differences as large as factors of two can be found even for nearby luminous galaxies.
For the satellite-dynamics galaxy sample of A16, A17 the adopted $r_e$ values are taken largely from
the ATLAS$^{3\mathrm{D}}$ survey \citep{cappellari+2013}, whereas for the X-ray sample we use the independently measured
values from B18.  These two lists correlate well (see Fig.~3 of B18) with 
little or no systematic offset, but the galaxy-to-galaxy
scatter is at the typical level of $\sim30-40$\%.  At $f_5 \simeq 0.5$, an uncertainty of
30\% in $r_e$ and thus in $M_5$ will produce an asymmetric external uncertainty of $(+0.12, -0.21)$ in $f_5$ from
this source alone.  
However, at $f_5 \simeq 0.8$ -- a level near the mean for our current data -- 
the resultant uncertainty range shrinks to $(+0.05, -0.11)$.

We should also consider 
the known systematic increase of $r_e$ with galaxy luminosity {\bf or mass}.
A recent compilation of optical measurements \citep{bender+2015} is well described
over the range $-16 \gtrsim M_V^t \gtrsim -24$ by 
\begin{equation}
    {\rm log}~ r_e {\rm (kpc)} = 0.411 - 0.256 (M_V^t + 20) \, .
    \label{normre}
\end{equation}
This relation is shown in the upper panel of Figure \ref{renorm}, along with the
measured data for our list of galaxies from Table 1.  
The slope is equivalent to a scaling very close to $r_e \sim L^{2/3}$, which is consistent
with the various versions of the fundamental plane for ETGs that relate
mass, $M/L$, and scale radius \citep[e.g.][]{burstein+1997,graham2019,chiosi+2019}.
As is evident from Fig.~\ref{renorm}, our independent measurements of $r_e$ versus luminosity are consistent with
this standard relation.   They are also closely consistent with
standard correlations of $r_e$ versus stellar \emph{mass} $M_{\star}$ \citep[see the references
cited above and also][for results from the GAMA sample]{lange+2015}, within 
the significant galaxy-to-galaxy scatter around these relations as noted above.
Within this scatter, our definitions of effective radius for both the simulated
galaxies and the observations are therefore consistent.

For each galaxy in our list, we can then calculate the normalized ratio $r_e/r_0$ where
$r_e$ is the measured radius and $r_0$ is the predicted value from Eqn.~\ref{normre}, 
given its luminosity $M_V^t$.  In the lower panel of Fig.~\ref{renorm}, we show $f_5$ versus this normalized
ratio.  No major correlation is evident.   A handful of objects with abnormally
large ratios ($r_e/r_0 > 3$) all have $f_5 > 0.8$.  These few cases may be
ones where the measured $r_e$ has been overestimated:  
if $r_e$ is too large, then $5 r_e$ will
be so large that the included halo volume will inevitably be more 
DM-dominated and thus $f_5$ will be larger than its true value.

The four cases with $f_5 \lesssim 0.4$ all have $(r_e/r_0) \lesssim 2.5$, which fall
within the main scatter seen in the upper panel of Fig.~\ref{renorm}.  
In other words, these low$-f_5$ cases do not seem to be due to overestimates
of $r_e$, which as noted above would tend to bias $f_5$ upward.  Lastly,
there are several objects with $(r_e/r_0) < 1$ and $f_5 \lesssim 0.7$.  
Though they are not strongly anomalous in the preceding figures of $f_5$ versus
mass, some of them could represent cases where $r_e$ (and thus $5 r_e$) 
is underestimated, leading to an overestimate of $(M_{bary}/M_{DM})$
and thus a decreased $f_5$.
Our tentative conclusion for this section is that at least some of
the outliers in the $f_5$ distributions, on both the high and low ends,
may be due to problems in the raw measurements of the galaxy effective
radii.

In Tables 1 and 2, the uncertainties listed are only the internal measurement
errors and do not include their external, method-dependent uncertainties that are much
harder to assess.  A better way to gauge the size of these effects may be what we
have done here, which is to compare two nearly independent methods for
estimating the DM mass fraction.  In the end, the mutual agreement is encouraging
despite some anomalous cases.

\section{Overview and Conclusions}

Seen broadly, the simulations as shown in Fig.~\ref{final} indicate that luminous galaxies can
realistically lie in the range $f_5 \simeq 0.6 - 0.9$ depending on the details of
their merger and growth histories.  High AGN activity, major mergers, and tidal
stripping of the outer halo may all contribute to lowering $f_5$, 
but in the present data few or no physically convincing cases fall below 
$f_5 \lesssim 0.5$, independent of mass.

Interestingly (and as also discussed in A17), the cases with $f_5 \lesssim 0.5$
predominantly fall in the relatively narrow mass range $M_{\star} \sim 0.6 - 3 \times 10^{11} M_{\odot}$.
Similarly, in the SDSS binned sample of \citet{wojtak_mamon2013}, 
$f_5$(min) =  0.63 is reached at $M_{\star} = 2.4 \times 10^{11} M_{\odot}$.
This mean mass is about 3 - 4 times \emph{higher} than 
the stellar mass $M_{\star} \simeq 6 \times 10^{10} M_{\odot}$ where the SHMR 
reaches its peak (Eq.~7 above), i.e.~where the global baryonic mass fraction is maximal.
As another comparison with theory, the predicted run of $f_5$ versus $M_{\star}$ from the
Illustris TNG simulations \citep{lovell+2018} goes through a shallow minimum of $f_5(min) \simeq 0.75$
at $M_{\star} \sim 2 \times 10^{10} M_{\odot}$, an order of magnitude lower mass than
we observe here.

Figure \ref{meanpoints} shows our final comparison of theory with data.  Here, the combined
list of 117 galaxies in our study is grouped into 13 mass bins of 9 galaxies each and the mean $f_5$ of
each bin is plotted versus stellar mass.  This distribution of mean points agrees well
with the mean trend from the SDSS satellite data, including the shallow dip 
near $M_{\star} \simeq 2 \times 10^{11} M_{\odot}$.  Comparing Figs.~\ref{final} and \ref{meanpoints} suggests that this dip is
not so much a downward shift of the \emph{median} $f_5$ at that mass, but rather
the presence of a distinctly larger proportion of low$-f_5$ galaxies in that critical
mass range.  This same dip is not evident in the simulations, and if real, it 
may present an interesting challenge for future modelling.

As one more extremely interesting comparison with 
a very different kind of observational data, we also show
the mean value of $f_5$ obtained by \citet{oguri+2014} from a sample
of 161 ETGs measured through the strong-lensing technique (see their Figure 7).  
The list of galaxies in their compilation covers the stellar mass 
range from $M_{\star} \simeq 2 \times 10^{10} M_{\odot}$ up
to $5 \times 10^{11} M_{\odot}$, overlapping well with our X-ray and satellite samples. 
For stellar masses they adopt the Salpeter IMF.  When these are adjusted by -0.3 dex to the
Chabrier/Kroupa (M/L) scale that we adopt here, their strong-lensing mean value
becomes $\langle f_5 \rangle = 0.86 \pm 0.06$, which sits higher by about 
$\Delta f_5 \simeq 0.05$ than the X-ray and satellite observations and the
centroid of the simulations.  Despite this small offset, 
the mutual agreement in mean $f_5$ among the very different techniques 
(strong-lensing, X-ray, and satellite dynamics) is encouraging.

Finally, the choice of the definition of stellar mass $M_*$ used for the simulations has a clear effect on comparison with the observed dark matter fractions $f_5$, as first
noted in Section 2 and Fig.~\ref{fig:sims}.  For the `normal' simulated galaxies, 
all four ways to
define $M_*$ place the simulated points somewhat below the observations in the
same mass range, but the choice of $M_*^2$ (= all stellar particles within the
virial radius, minus substructure) gives the best overall agreement.  It should
also be noted, however, that in this lower-mass
regime ($M_* \lesssim 10^{11} M_{\odot}$) on the observational side, 
the X-ray signals are the weakest and the mass estimates subject to 
uncertainties that are hard to assess.
On the high-mass end (the `centrals' or BCGs) the best agreement comes with
the choice of $M_*^1$ (predicting stellar mass from total halo mass by
inversion of Eq. 7).  Though the mean $f_5$ level for the centrals does not change much for any of
the four definitions of $M_*$, the range of estimated stellar masses 
matches the observations best for $M_*^1$.  It is perhaps also worth
noting that the relative numbers of low$-f_5$ outliers are fairly
similar for the different definitions.

\section{Some Future Prospects}

To first order, we are encouraged by the general agreement between simulations
and observations (quantified in Tables 3 and 4).  For the entire mass range
$M_* > 10^{10} M_{\odot}$, the mean $\langle f_5 \rangle$ remains in the range 
$0.7 - 0.9$ with no large systematic trend with either galaxy mass or
morphology.  At a finer level of detail, interesting
features of the distribution remain that may be due to measurement scatter, 
problems with the analytical methods (both X-ray and satellites), or
genuine differences in galaxy histories, that still need to be
sorted out.

Ultimately, the choice of $5r_e$ as a fiducial radius for evaluating DM fractions
remains, at the very least, a little arbitrary.  In the longer term, more information and a
better understanding of formation histories should come from measurements of
the \emph{radial profile} of $f_{DM}(r)$ for any given galaxy, from its inner halo 
out to the observational limits that the data permit.  Different fiducial radii
such as $1, 2, 3,$ or $5r_e$ have been used by some other authors
\citep[e.g.][A16, A17]{deason_etal2012,wojtak_mamon2013}, but more
continuous radial profiles are within reach.  These goals will be addressed
in future papers,
but a brief look at their potential is illustrated in Figure \ref{expanded}.
Here, the more general DM mass fraction within 3D radius $R$ is plotted versus
radius in units of the half-mass radius $R_{1/2}$ as defined previously, for all normal galaxies from the Magneticum simulations.
The simulated systems define the family of (grey) curves $f_{DM}(R)$ that in most cases
rise steeply from $\sim 1 - 5 R_{1/2}$ and then flatten off once we are so
far out in the halo that the enclosed baryonic mass is no longer changing.
How far to the left or right each curve falls is a marker of its halo central
concentration and the particular evolutionary history it has experienced.
Fiducial curves are also shown for various NFW-type halos from \cite{courteau+2015}.
These include curves for a standard ETG as well as a late-type galaxy (LTG),
along with two additional curves for these models in which the DM halo is
adiabatically contracted (dotted and dash-dotted lines).  The standard (non-contracted)
model halos clearly make a better match to the curves of growth for the simulated 
galaxies, even though several simulated halos show clear signs of at least some contraction. This spread in dark matter radial distribution, with most galaxies being close to NFW-like or (slightly) contracted, is in good agreement with results from the IllustrisTNG simulations, as shown by \citet{wang+2020}, and also with the strong-lensing results from \citet{oguri+2014} who find that the dark matter halos of the galaxies in their sample are closer to NFW profiles than contracted profiles.

The cases with lower than average dark matter fractions ($f_5 < 0.6$ using stellar mass
$M_*^2$) are plotted as blue curves.  Interestingly, all of these low$-f_5$ galaxies from
the Magneticum simulations have centrally \emph{expanded} dark matter halos as evidenced by the much flatter inner dark matter fraction slopes; see Fig.~\ref{expanded}. We find these expanded halos at all mass ranges, different from \citet{wang+2020} who find these halos only at the low mass end of $M_* < 10^{11}M_\odot$, but their expanded halos also have rather low dark matter fractions at least within one halfmass radius. A potential explanation for the expansion of halos is strong outflows from stellar feedback (e.g \cite{governato:2012} using Zoom simulations with Gasoline, and \cite{dutton+2016} using the NIHAO simulations) or AGN feedback \citep[][using HorizonAGN]{peirani:2017}, and feedback is most likely also the reason for the expanded halos in both Magneticum and IllustrisTNG. However, analysing this interplay between stellar and dark component leading to contraction and expansion of dark matter halos is beyond the scope of this paper and will be addressed in a followup study.

Since contracted and expanded halos are clear signs of different dominant formation scenarios, it would be extremely helpful in deciphering individual histories to measure the amount of contraction or expansion in detail. The contracted halos raise the DM fraction dramatically at small radii but have relatively less effect at several effective radii beyond most of the baryonic matter.  Contrarily,  the expanded halos differ strongly at larger radii. Measuring radial dark matter profiles 
(that is, the curves of growth as seen in Fig.~\ref{expanded}) from observations would be invaluable.

The observations (magenta dots in Fig.~\ref{expanded}), obviously, sample the theoretical curves at only one radius.  A re-analysis of the data for a series of radii, not just $5 r_e$ (both
for the X-ray and satellite techniques), would generate 
these curves of growth directly from the observations and would lead to 
a more informative confrontation with the simulations.
The current list of observations at $5 r_e$ is not sufficient for strong conclusions about which model
curves might be better, but it is potentially
interesting to note that the median of the observed distribution indicates contracted halos instead of classical standard halos. This again points to the need to develop more complete curve of growth  particularly at smaller radii.  Similarly, for those halos with especially low $f_{DM}$ we may 
be able to determine whether or not they are expanded halos.

\section{Summary}

We have used measurements of the X-ray gas that can be found in individual large galaxies for
a new assessment of the mass fraction $f_5 = f_{DM}(5r_e)$ of dark matter
within the fiducial radius of $5 r_e$.  This mass measurement technique is nearly independent of the  more widely used satellite kinematics methods.  The results for our sample of 102 galaxies are compared
directly with theoretical predictions from the Magneticum/Pathfinder
suite of simulations for both normal galaxies and centrally dominant
systems (BCGs and BGGs). A summary of our findings is as follows:

\begin{enumerate}
\item Over the mass range of our sample of 102 galaxies
($10^{10} M_{\odot} \lesssim M_{\star} \lesssim 10^{12} M_{\odot}$), we find that $f_{DM}$ stays at a median level near $f_5 \sim  0.85$ and rms scatter $\pm 0.15$, nearly independent
of mass.  
    \item The observed distribution of $f_5$ shows substantial
    agreement with the simulations in the mean (high) level of $f_5$ and the
    galaxy-to-galaxy scatter (cf. Fig.~\ref{final}).  The observed data, however,
    show some individual galaxies scattering to both higher and much lower $f_5$ values
    that are only rarely reached by the simulated systems.  Many, though not all, of these
    extreme cases may be real.  In general, the galaxy-to-galaxy spread in halo dark matter fraction 
    points to the diverse formation pathways including feedback, 
    outflows and inflows, and mergers of different mass fractions that exist for galaxies 
    in this mass range.  This diversity clearly emphasizes the importance of 
    further combined studies from simulations and observations.
    \item In our X-ray sample, the disk-type galaxies (S0/SA0/SB0) have a significantly
    higher dark-matter fraction (median $f_5 \simeq 0.9$) than do the E-types ($f_5 \simeq 0.8$).
    If physically real, this difference is, perhaps, an indicator of their quieter and 
    earlier history of growth by mergers and satellite accretions, less halo expansion
    due to feedback, and very sparse stellar halo components.
\item Though with differences in detail, the overall pattern of $f_5$ from the X-ray
measurements generally agrees well with recent measurements from satellite dynamics (A17).  The median and mean $f_5$ levels from both X-ray and satellite methods
    also agree to within their uncertainties with the
    estimate of $\langle f_5 \rangle = 0.86 \pm 0.06$ obtained from strong lensing,
    and from a different sample of galaxies.
    To first order, there is now an encouraging concordance among three quite different
    methods for measuring DM mass fraction in the outer halos of large galaxies:  satellite dynamics, 
    strong lensing, and X-ray gas profiles.
    \item For the central giant galaxies (BCGs or BGGs), there is good agreement
    between the 44 observed cases and the predicted level from the simulations.
    This result emphasizes
    the strong dark matter dominance within the BCGs.
    \item In all subgroups of the data (E-type galaxies, disk types, BCG/BGG types, X-ray or
    satellite methodology), the observed internal scatter at any
    mass is $\sigma(f_5) \simeq 0.15$, which is larger than the typical measurement uncertainty
    for most individual galaxies.  For comparison, the internal scatter of the simulated systems
    is $\sigma(f_5) \lesssim 0.1$, which appears to be primarily because the simulations do
    not predict the same fractions of galaxies at extremely high or low $f_5$ (outliers).
    \item Further studies to generate more observational DM mass fraction \emph{profiles}
    covering a much bigger range in radii will be crucial for making
    deeper connections with the simulations.
\end{enumerate}

\section*{Acknowledgements}
This research has made use of data obtained from the Chandra Data Archive 
and the Chandra Source Catalog, and software provided by the Chandra X-ray Center 
(CXC) in the application packages CIAO, ChIPS, and Sherpa. 
We thank all the staff members involved in the Chandra project. 
We are grateful to Mike Hudson and Ian Roberts for their advice, and to Gary Mamon
for transmitting the SDSS comparison data shown in the figures.  
WEH acknowledges the financial support of NSERC.
The {\it Magneticum} Pathfinder simulations were performed at the Leibniz-Rechenzentrum with CPU
time assigned to the Project {\it pr86re} and supported by the Deutsche
Forschungsgemeinschaft (DFG, German Research Foundation) under
Germany's Excellence Strategy - EXC-2094 - 390783311.
Lastly, we thank the anonymous referee for helpful comments and suggestions.

%\makeatletter\@chicagotrue\makeatother

\bibliographystyle{apj}
%\bibliography{xray}
\label{lastpage}

\end{document}